\documentclass[aps,nofootinbib,twocolumn,groupedaddress,floatfix,eqsecnum]{revtex4}
\usepackage[pdftex]{graphicx}
\usepackage{rotating}
\usepackage[mathscr]{eucal}
\usepackage{amsmath,amssymb,bm,graphics}
\usepackage[letterpaper=true]{hyperref}

\def\coeff#1#2{{\textstyle \frac{#1}{#2}}}
\let\mathbf=\bm

\begin{document}
\def\Nfour{\mathcal N\,{=}\,4}
\def\Ntwo{\mathcal N\,{=}\,2}
\def\Nc{N_{\rm c}}
\def\Nf{N_{\rm f}}
\def\x{\mathbf x}
\def\q{\mathbf q}
\def\f{\mathbf f}
\def\v{\mathbf v}
\def\C{\mathcal C}
\def\w{\omega}
\def\vs{v_{\rm s}}
\def\S{\mathcal S}
\def\half{{\textstyle \frac 12}}
\def\twothirds{{\textstyle \frac 23}}
\def\third{{\textstyle \frac 13}}
\def\t{\mathbf{t}}
\def\T{\mathcal {T}}
\def\O{\mathcal{O}}
\def\E{\mathcal{E}}
\def\p{\mathcal{P}}
\def\H{\mathcal{H}}
\def\uh{u_h}
\def\R{\ell}
\def\Ro{\psi}
\def\del{\nabla}
\def\nn{\nonumber}
\def\K{\mathcal K}
\def\inf{.pdfilon}
\def\cs{c_{\rm s}}
\def\A{\mathcal{A}}
\def\e{{e}}
\def\r{{\xi}}
\def\x{{\mathbf x}}
\def\w{{w}}
\def\rr{{\xi}}
\def\uo{{u_*}}
\def\to{{t_*}}			
\def\u{{\mathcal U}}
\def\G{\mathcal{G}}
\def\Deltax{\Delta x_{\rm max}}
\def\xendpt{x_{\rm endpt}}
\def\xendpt{\mathcal X}

\title
{Light quark energy loss in strongly-coupled \boldmath
$\Nfour$ supersymmetric Yang-Mills plasma }

\author{Paul~M.~Chesler\footnotemark}
\author{Kristan~Jensen\footnotemark}
\author{Andreas~Karch\footnotemark}
\author{Laurence~G.~Yaffe\footnotemark}

\affiliation
    {Department of Physics, University of Washington, Seattle, WA 98195, USA}

\date{\today}

\begin{abstract}
We compute the penetration depth of a light quark moving through
a large $\Nc$, strongly coupled $\Nfour$ supersymmetric
Yang-Mills plasma using gauge/gravity duality and a combination
of analytic and numerical techniques.
We find that the maximum distance
a quark with energy $E$ can travel through a plasma is given by
$\Deltax(E) = (\C/T) \, (E/T \sqrt\lambda)^{{1}/{3}}$
with $\C \approx 0.5$.
\end{abstract}


\maketitle
\iftrue
\def\thefootnote{\fnsymbol{footnote}}
\footnotetext[1]{Email: \tt pchesler@u.washington.edu}
\footnotetext[2]{Email: \tt kristanj@u.washington.edu}
\footnotetext[3]{Email: \tt karch@phys.washington.edu}
\footnotetext[4]{Email: \tt yaffe@phys.washington.edu}
\def\thefootnote{\arabic{footnote}}
\fi

\section{Introduction}

The discovery that the quark-gluon plasma produced
at RHIC behaves as a nearly ideal fluid \cite{Shuryak,Shuryak:2004cy}
has prompted much
interest into the dynamics of strongly coupled plasmas.
Hard partons produced in the early stages of heavy ion collisions
can traverse the resulting fireball and deposit their energy and
momentum into the medium.  Analysis of particle correlations
in produced jets can provide useful information about the dynamics
of the plasma including the rates of energy loss and momentum broadening
\cite{Leitch:2006ex,CasalderreySolana:2006sq},
as well as the speed
and attenuation length of sound waves \cite{CasalderreySolana:2006sq}.

Gauge/gravity duality
\cite{Maldacena:1997re, Witten:1998qj,Gubser:1998bc,Aharony:1999ti}
is a useful tool for the study of dynamics of
strongly coupled non-Abelian plasmas.  Although
no gravitational dual to QCD in known, gauge/gravity duality
has provided much insight into the dynamics of various theories
which share many qualitative properties with QCD.  The most widely
studied example is that of strongly coupled $\Nfour$
supersymmetric Yang-Mills theory (SYM).
The deconfined plasma phases of
QCD and SYM share many properties.
For example, both theories describe non-Abelian plasmas
with Debye screening, finite spatial correlation lengths,
and long distance dynamics described by neutral fluid hydrodynamics.
When both theories are weakly coupled, appropriate comparisons
of a variety of observables show rather good agreement
\cite{CaronHuot:2006te,Huot:2006ys,Chesler:2006gr}.
This success, combined with the lack of alternative techniques
for studying dynamical properties of QCD at temperatures where
the plasma is strongly coupled, has motivated much interest in using
strongly coupled $\Nfour$ SYM plasma as a model for QCD plasma
at temperatures of a few times $\Lambda_{\rm QCD}$
(or $1.5 \, T_{\rm c} \lesssim T \lesssim 4 \, T_{\rm c}$).
(See, for example, Refs.~\cite{
Kovtun:2004de,
Herzog:2006gh,
Casalderrey-Solana:2006rq,
Gubser:2006nz,
Gubser:2006qh,
Liu:2006ug,
Shuryak:2006ii,
Lin:2006rf,
CasalderreySolana:2007qw,
Lin:2007pv,
Lin:2007fa,
Hatta:2007cs,
Dusling:2008tg,
Liu:2008tz},
and references therein.)
At least for some quantities, this is quite successful.
In particular, the value of the shear viscosity
to entropy density ratio \cite{Policastro:2001yc,Kovtun:2004de},
$\eta/s = 1/4 \pi$, in strongly coupled SYM 
is in rather good agreement with estimates which emerge from
hydrodynamic modeling of heavy ion collisions \cite{Luzum:2008cw}.

In the limit of large $N_c$ and large 't Hooft coupling
$\lambda \equiv g^2 \Nc$,
the gravitational dual to $\Nfour$ SYM is described by classical supergravity
on the ten dimensional $AdS_5 \times S^5$ geometry \cite{Maldacena:1997re}.
Studying the theory at finite temperature
corresponds to adding a black hole to the geometry \cite{Witten:1998qj}.
The corresponding AdS-Schwarzschild (AdS-BH)
metric is given below in Eq.~(\ref{metric}).
Fundamental representation
quarks added to the $\Nfour$ theory theory are dual to open strings moving
in the $10d$ geometry.
In the limit of large $\lambda$, where the string action and the
energy both scale like $\sqrt{\lambda}$, quantum fluctuations in the
string worldsheet are suppressed and the dynamics of strings may be described
by the classical equations of motion which follow from the Nambu-Goto action.

The dynamics of strings corresponding to heavy quarks have been intensely
studied by many authors.
The energy loss rate for heavy quarks moving through
a SYM plasma has been studied
in Refs~\cite{Herzog:2006gh,Casalderrey-Solana:2006rq,Gubser:2006bz,Chernicoff:2008sa},
and the wake produced by a moving heavy quark was computed in
Refs.~\cite{Chesler:2007an,Chesler:2007sv,Gubser:2007xz,Gubser:2007ga}.

Analogous studies for light quarks have yet to be completed.
In Ref.~\cite{Chesler:2008wd} the charge density of massless
quarks moving through a SYM plasma was studied,
and it was shown that there are string states which are dual to long-lived
excitations (\textit{i.e.}, quasi-particles) in the field theory.
In particular, the charge density of highly energetic light quarks
can remain localized for an arbitrarily long time,
and can propagate arbitrarily far before spreading out and thermalizing.
In Ref.~\cite{Gubser:2008as}, an
attempt was made to estimate the penetration depth of a gluon moving through
a strongly coupled SYM plasma.  The results of Ref.~\cite{Gubser:2008as}
were obtained by assuming that the endpoint of a (folded)
string follows a light-like geodesic in the AdS-BH geometry;
full solutions to the string equations of motion were not obtained.
The authors of
Ref.~\cite{Gubser:2008as} tried to roughly characterize the relationship
between the string's energy and momentum and the parameters of the geodesic,
and suggested that the maximum distance
a gluon of energy $E$ can go before thermalizing should
scale as $\Deltax \sim E^{{1}/{3}}$.
The same scaling relation has also been 
discussed for $R$-current jets in Ref.~\cite{Hatta:2008tx}.

Although the estimates made in Ref.~\cite{Gubser:2008as}
are generally plausible, we believe that it is clearly desirable
to perform a quantitative, controlled study of the penetration
distance of light quarks (or gluons) in a strongly coupled plasma.
This is a key aim of this paper.

It should be emphasized that we are concerned with studying the
propagation through the plasma of energetic excitations which
resemble well-collimated quark jets.
The open string configurations we consider may be regarded
as providing a dual description of dressed quarks, with high energy,
moving through a non-Abelian plasma.
We are not studying the result of a local current operator
acting directly on the strongly coupled $\Nfour$ SYM plasma.
(See, however, Ref.~\cite{Hofman:2008ar}.)
Our motivation is similar to that of Ref.~\cite{Liu:2006ug}, in which
weak coupling physics in asymptotically free QCD is envisioned
as producing a high energy excitation, whose propagation through
the plasma is then modeled by studying the behavior of the same
type of excitation in a strongly coupled $\Nfour$ SYM plasma.

The energy loss rate for a heavy quark depends only on the quark's velocity,
the value of the 't Hooft coupling $\lambda$,
and the temperature of the plasma through which the quark is moving
\cite{Herzog:2006gh}.
In other words,
for very heavy quarks which slowly decelerate,
the velocity is the only aspect of their initial conditions which
influences the energy loss rate.
This turns out not to be the case for light quarks.
Initial conditions for a classical string involve two free functions:
the initial string profile and its time derivative.
As we discuss in detail below, the instantaneous energy loss rate
of a light quark depends strongly, in general, on the precise choice
of these initial functions.
In the dual field theory,
this reflects the fact that any complete specification of an initial
state containing an energetic quark must also involve a characterization
of the gauge field configuration.
In the perturbative regime, one can easily see that the interactions
of heavy particles with a gauge field are spin-independent
(up to $1/M$ corrections), but interactions of relativistic particles are
spin-dependent even at leading order.
So it is perhaps unsurprising that the energy loss of a light projectile
also depends on the configuration of the gluonic cloud surrounding the
projectile in a non-universal fashion.

One quantity which is rather insensitive to the precise initial
conditions of the string is the maximum distance $\Deltax(E)$
which a quark with initial energy $E$ can travel.
It should be emphasized that we are considering effectively
on-shell quarks which can travel a large distance $\Delta x$
before thermalizing.
The maximum penetration depth $\Deltax$ grows without bound
as the energy $E$ increases.

\begin{figure}
\includegraphics[scale=0.37]{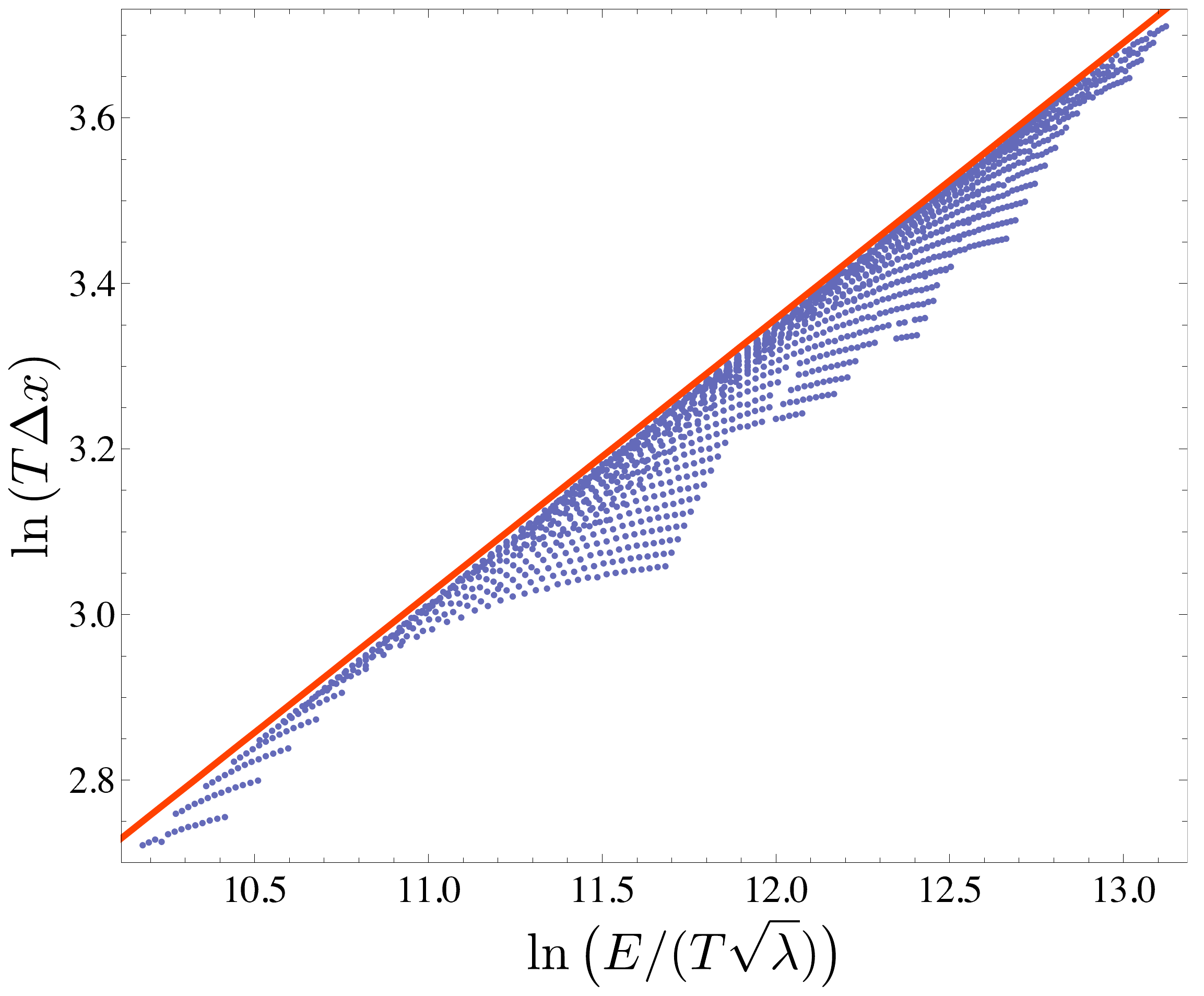}
\caption
  {
  \label{allData}
  A log-log plot of the quark stopping distance $\Delta x$
  as a function of total quark energy $E$
  for many falling strings with
  initial conditions of the form shown in Eq.~(\ref{IC}).
  All data points fall below the red line given by
  $\Delta x = ({0.526}/{T})\big( {E}/{T\sqrt{\lambda}} \big)^{1/3}$.
}

\end{figure}

We numerically compute the penetration depth $\Delta x$ for many
different sets of string initial conditions,
and find that the maximum penetration depth does
indeed scale like $E^{{1}/{3}}$.
Our results are illustrated in Fig.~\ref{allData}\,, where the logarithm
of the penetration depth is plotted as a function of the logarithm of
the initial quark energy for many different sets of initial conditions.
As is evident from the figure, the penetration depth of a light quark
is bounded by a curve $\Deltax = \text{const.} \times E^{1/{3}}$.

We also demonstrate the scaling relation $\Deltax \sim E^{1/{3}}$
analytically.  As discussed in Ref.~\cite{Chesler:2008wd}, strings
which correspond to long-lived massless quarks are approximately
null strings.  A strictly null string is one whose worldsheet metric is
everywhere degenerate.  The qualitative origin of this connection is easy
to understand.
Strings which correspond to light quarks fall into the event horizon.
As they fall they become more and more light-like and hence closer and
closer to a null configuration as time progresses.  The profile of the
null string is almost independent of the initial conditions used to
create the string --- for the quasiparticle excitations studied in this
paper, the corresponding null strings are specified by two numbers only,
an initial inclination and radial depth.
By analyzing strings corresponding to light quarks as small perturbations
away from null string configurations, we show that the total distance a
quark can travel must be bounded by a maximum distance
$\Deltax = (\C/T) (E/T \sqrt\lambda)^{1/{3}}$
for some $O(1)$ constant $\C$.
We numerically confirm that strings corresponding to
long-lived light quarks are, in fact, close to being null,
and obtain an estimate of the constant $C$.

Although the endpoint motion of our string
solutions is well approximated by appropriate light-like geodesics,
consistent with the discussion of Ref.~\cite{Chesler:2008wd},
we find that the relationship between the parameters of the geodesic
and the string profile and energy is more complicated (and rather
different) than the surmises presented in Ref.~\cite{Gubser:2008as}.
This will be discussed further in Section~\ref{discussion}\,.

In addition to studying the penetration depth,
we also examine the instantaneous rate of energy loss, $dE/dt$.
For light quarks the energy loss rate shows non-universal
features and is sensitive to initial conditions.
For the states we study, we find that it typically {\em increases\/}
with time during the period when the dressed quark is a
well-defined quasiparticle
and sharply peaks during the final thermalization phase.  
In other words, the thermalization of light quarks in strongly 
coupled SYM ends with an explosive burst 
of energy.  This late time behavior is universal 
and independent of initial conditions.  The fact that the 
light quark energy loss rate can increase with time is qualitatively 
different from the behavior of heavy particles \cite{Herzog:2006gh},
whose energy loss rate monotonically decreases.

An outline of our paper is as follows.
We define our conventions in Section~\ref{conventions}\,.
In Section~\ref{lightquark}\,, we discuss some of the subtleties
involved in defining the light quark energy loss rate
and penetration depth, and spell out the relevant identifications
between $5d$ gravitational and $4d$ field theory quantities.
In Section~\ref{fallingstrings}\,,
we discuss the dynamics of strings both from analytical and numerical perspectives.
Implications of our results, and connections with other related work,
are discussed in Section~\ref{discussion}\,,
which is followed by a brief conclusion.

\section{Conventions}
\label{conventions}

Five dimensional AdS coordinates will be denoted by $X_M$, while
four dimensional Minkowski coordinates are denoted  by $x_\mu$.
Worldsheet coordinates will be denoted as
$\sigma^a$ with $a = 0,1$.
The timelike world sheet coordinate is $\tau \equiv \sigma^0$,
while the spatial coordinate is $\sigma \equiv \sigma^1$.
When discussing the dynamics of a single string endpoint,
we will use $\sigma^*$ to denote the value of $\sigma$ at the endpoint.

We choose coordinates such that the metric of the AdS-Schwarzschild
(AdS-BH) geometry is
\begin{equation}
\label{metric}
    ds^2 = \frac{L^2}{u^2}
    \left [-f(u) \, dt^2 + d \x^2 + \frac{du^2}{f(u)} \right ] ,
\end{equation}
where $f(u) \equiv 1-(u/u_h)^4$ and $L$ is the AdS curvature radius.
The coordinate $u$ is an inverse radial coordinate;
the boundary of the AdS-BH spacetime is at $u = 0$
and the event horizon is located at $u=u_h$,
with $T \equiv (\pi u_h)^{-1}$ the temperature of the equilibrium SYM plasma.

\section{Light quarks and gauge/gravity duality}
\label{lightquark}

\begin{figure}
\includegraphics[scale=0.45]{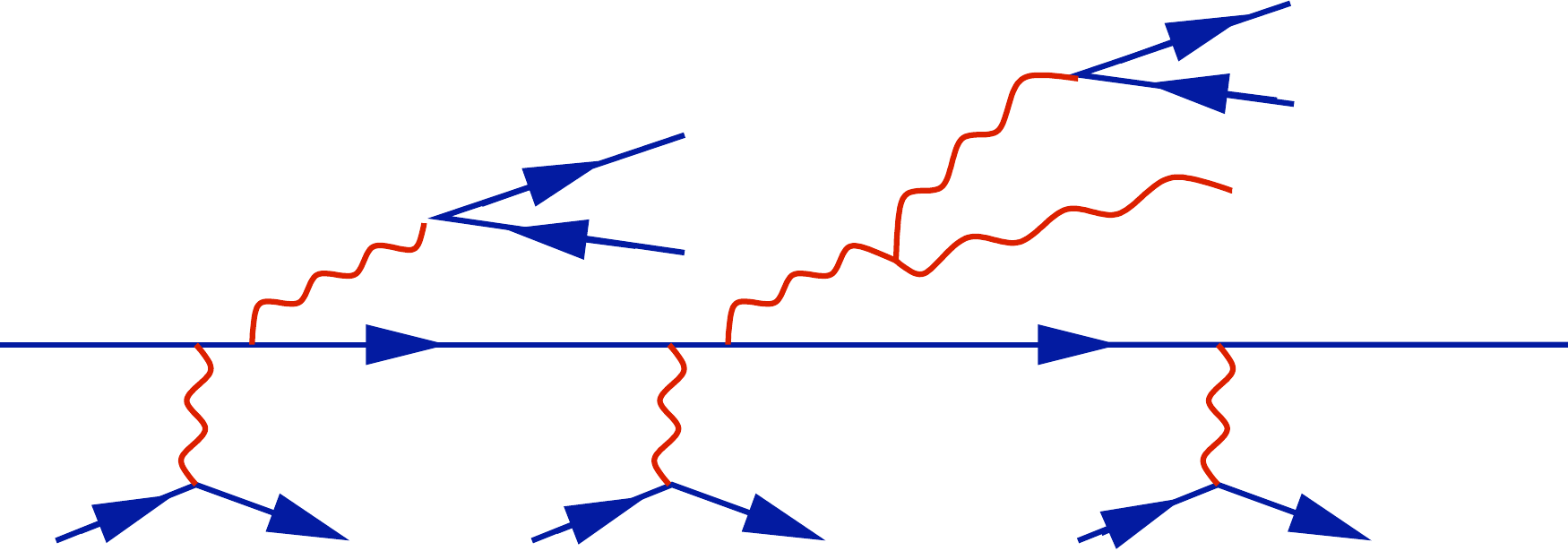}
\vspace*{10pt}

\includegraphics[scale=0.45]{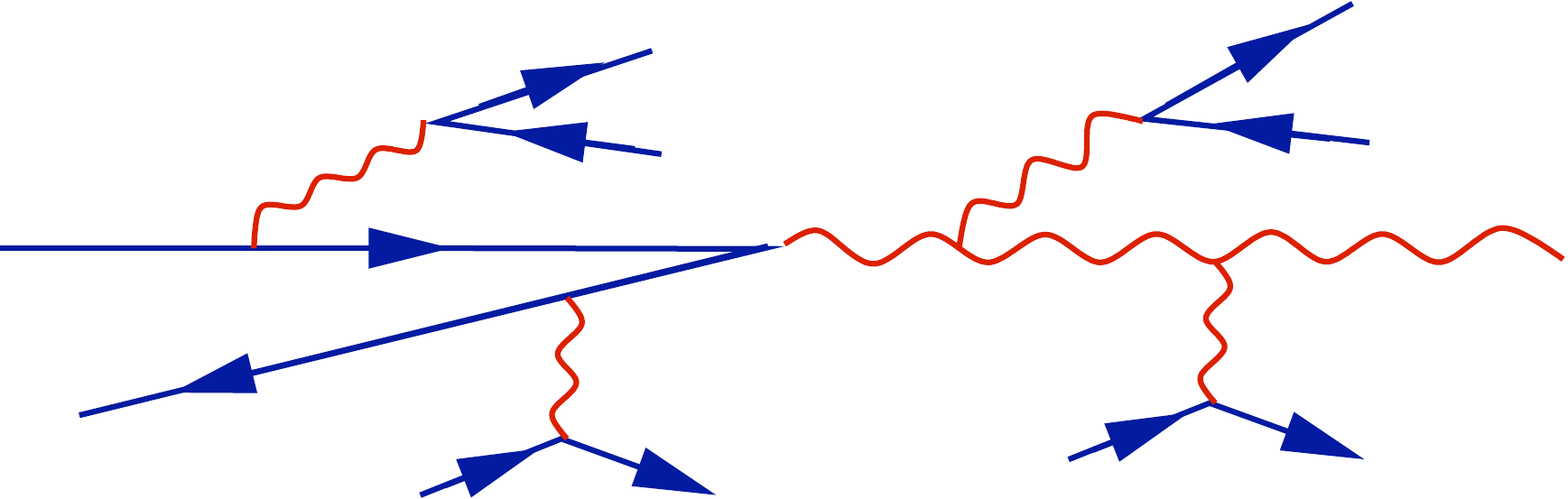}
\caption{\label{shower}
    Examples of perturbative diagrams contributing to the
    energy loss of a quark.
    One may regard time as running to the right.
    An energetic quark can scatter,
    emit gluons (which themselves may radiate or split into $q \bar q$ pairs),
    or annihilate with an antiquark in the medium.
    However, the total baryon number of the state remains constant.
    }
\end{figure}

Energetic quarks moving through a plasma are quasi-particles ---
they have a finite lifetime which can be long compared to the inverse
of their energy.
Some care is needed in defining the light quark
penetration depth and the instantaneous energy loss rate.
Fig.~\ref{shower} shows some typical perturbative diagrams contributing
to the energy loss rate of a quark.
A energetic quark, scattering off excitations in the medium,
can emit gluons which may subsequently
split into further gluons or quark-antiquark pairs.
The energetic quark may also annihilate with an antiquark in the medium.
A natural question to consider when looking at
Fig.~\ref{shower} is which quark should one follow when computing
the penetration depth?
Once a quark has emitted a $q \bar q$ pair,
or annihilated with an antiquark,
it becomes ambiguous which quark was the original one.
This issue is cleanly avoided if one focuses attention not on some
(ill-defined) ``bare quark'', but rather on the baryon density of the
entire dressed excitation.

In QCD, or $\Nfour$ SYM coupled to a fundamental representation
$\Ntwo$ hypermultiplet,
there is a conserved current which we will
call $J^{\mu}_{\rm baryon}$.
Even though $q \bar q$ pairs can be produced by an energetic quark
traversing the plasma, conservation of
$J^{\mu}_{\rm baryon}$ implies that the total
baryon number of the excitation will remain constant.
The baryon density of an energetic excitation can remain
highly localized for a long period of time.
It is the collective excitation with localized baryon density
which we will refer to as a dressed quark,
or for the sake of brevity, simply as a quark.

To evaluate the penetration depth of a quark, we will use the centroid
of the baryon density,
\begin{equation}
\bar {\mathbf x}(t)  \equiv
\frac{\int d^3 x \> \mathbf x \, \rho(t,\mathbf x)}
    {\int d^3 x \> \rho(t,\mathbf x)} \,.
\end{equation}
with $\rho \equiv J^0_{\rm baryon}$.
The centroid $\bar {\mathbf x} (t) $ gives a natural
measure of where the quark is located at time $t$.
At late times, when the quark has lost nearly all its energy and becomes
thermalized,
the dynamics of the baryon density will be governed by hydrodynamics.
In particular, at late times the baryon density must satisfy the
diffusion equation
\begin{equation}
\left(\partial_0 - D \del^2 \right ) \rho = 0 \,,
\end{equation}
where $D$ is the baryon number diffusion constant.
When the diffusion equation is applicable, it is easy to see that
the centroid ceases to move,
$ {d{\bar {\mathbf x}} }/{dt} = 0$.
To define the penetration depth, we imagine measuring
$\bar x(t)$ at some early time $\to$.
We then define the penetration depth $\Delta x$ in the obvious manner as
\begin{equation}
\label{Deltax1}
\Delta x \equiv |\bar {\mathbf x} (\infty) -\bar {\mathbf x} (\to)| \,.
\end{equation}

On the gravitational side of the gauge/gravity correspondence,
the addition of a $\Ntwo$ hypermultiplet to the $\Nfour$ SYM theory
is accomplished by adding a $D7$ brane to the $10d$ geometry
\cite{Karch:2002sh}.
The $D7$ brane
fills a volume of the AdS-BH geometry which extends from the boundary
at $u = 0$ down to 
maximal radial coordinate $u_m$, and wraps an $S^3$ of the $S^5$.
The bare mass $M$ of the hypermultiplet is proportional to
$1/u_m$ \cite{Herzog:2006gh},
so for massless quarks the
$D7$ brane fills all of the five-dimensional AdS-BH geometry.
Open strings which end on the $D7$ brane represent \textit{dressed}
$q \bar q$ pairs in the field theory.
In the $5d$ geometry these strings can fall unimpeded
toward the event horizon until their endpoints reach the radial
coordinate $u_m$ where the $D7$ brane ends.%
\footnote
    {%
    One should bear in mind that even when the radial position
    of the string endpoints lies closer to the boundary than $u_m$,
    the string endpoints are nevertheless attached to the
    $D7$ brane, albeit in the full $10d$ space.
    The embedding of the $D7$ brane is determined dynamically by
    minimizing the $D7$ worldvolume. In general, this means that the
    $D7$ brane wraps a 3-sphere inside the $S^5$ of the AdS-BH${} \times S^5$
    background geometry.
    This 3-sphere varies in a non-trivial way as a function
    of the radial coordinate of the AdS-BH geometry.
    For a hypermultiplet with non-zero mass,
    the string endpoints must move on the internal $S^5$
    as they fall down in the AdS-BH geometry,
    so that the string endpoints remain on the $D7$ brane.
    But for massless hypermultiplets, the corresponding
    $D7$ embedding is a simple product space, AdS-BH $\times S^3$.
    In this case, it is consistent to have the entire string sit at
    a fixed point on the $S^5$ while it falls in the AdS-BH background.
    Any additional motion of the string in the internal space will only
    add to the energy of the string without affecting its stopping distance
    and so will be of no interest for us ---
    we want to find strings which carry a minimal amount of energy
    for a given stopping distance.
    In the large $N_c$ limit, one can ignore the backreaction
    of the $D7$ brane on the background geometry and the backreaction
    of the string on the $D7$,
    as well as potential instabilities involving 
    string breaking or dissolving into the D-brane.
    These issues are discussed further in Section~\ref{discussion}\,.
    }
For sufficiently light or massless quarks, $u_m>u_h$ and open string
endpoints can fall into the horizon.%
\footnote
    {%
    Strictly speaking,
    in the coordinate system we are using no portion of the string
    crosses the horizon at any finite value of time.
    Due to the gravitational redshift,
    the rate of fall $du/dt$ decreases exponentially 
    as one approaches the horizon.
    Nevertheless, it is natural to speak of the string endpoint
    falling ``into'' or ``reaching''
    the horizon when $u - \uh \ll \uh$.
    }

\begin{figure}
\includegraphics[scale=0.35]{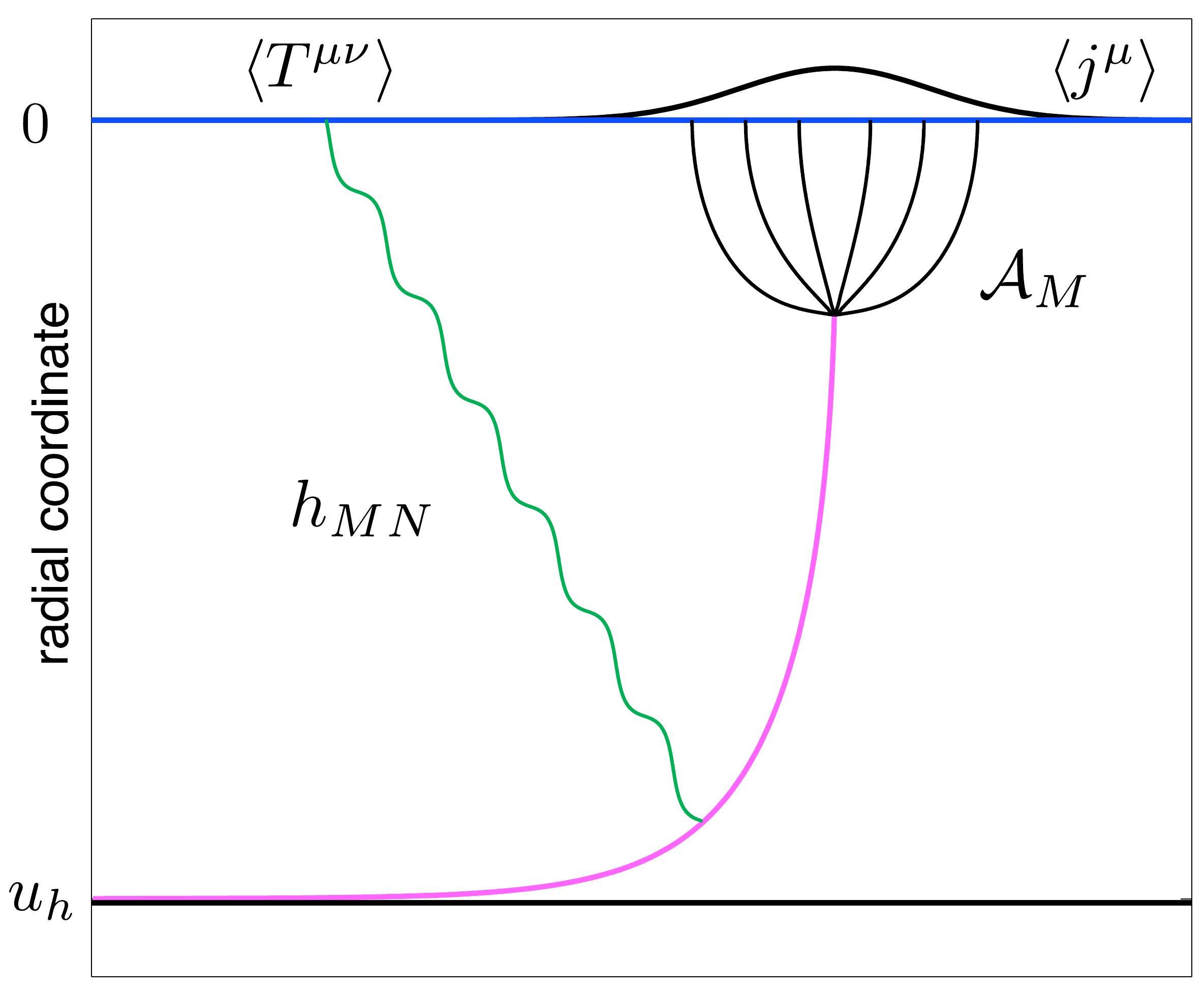}
\caption
    {\label{bulk2boundary}
    A cartoon of the bulk-to-boundary problem at finite temperature.
    The endpoints of strings are charged under a $U(1)$ gauge field
    $\mathcal A_M$ which lives on the $D7$ brane which fills the AdS-BH
    geometry.
    The boundary of the geometry,
    located at radial coordinate $u = 0$, behaves like a perfect conductor.
    Consequently, the string endpoints induce a 
    mirror current density $j^{\mu}$ on the boundary.
    Via gauge/gravity duality, the induced mirror current density has the
    interpretation of minus the baryon current density of a quark.  Similarly, the
    presence of the string induces a perturbation $h_{MN}$ in the metric
    of the bulk geometry.  The behavior of the metric perturbation near
    the boundary encodes the information contained in the perturbation
    to the SYM stress-energy tensor caused by the presence of the jet.
    }
\end{figure}

The endpoints of strings
are charged under a $U(1)$ gauge field $\mathcal A_{M}$ which resides on the
$D7$ brane.  The boundary of the $5d$ geometry, which is where the field theory
lives, behaves as an ideal electromagnetic conductor \cite{Kovtun:2005ev}
and hence the presence of the string endpoints, which source the
$D7$ gauge field $\mathcal A_{M}$,
induce an \textit{image current density} $J^{\mu}_{\rm baryon}$ on the boundary.
This is illustrated schematically by the cartoon in Fig.~\ref{bulk2boundary}\,.
Via the standard gauge/gravity dictionary
\cite{Maldacena:1997re,Witten:1998qj,Aharony:1999ti,Gubser:1998bc,Karch:2002sh},
the induced current density corresponding to each string
endpoint has a field theory interpretation as minus the 
baryon current density of a dressed quark.%
\footnote
  {%
  The fact that the induced mirror current density is minus the physical
  baryon current density is easy to understand.
  The baryon current density is given by the variation of the on-shell
  electromagnetic action with respect to the boundary value of the gauge
  field $\mathcal A_{M}$.  The on-shell $5d$ electromagnetic action evaluates 
  to a $4d$ surface integral, evaluated at the boundary with an outward
  pointing normal.
  In contrast, the image current density induced on the boundary
  can be obtained by integrating the $5d$ Maxwell equations
  over a Gaussian pillbox which encloses the boundary.
  The resulting surface integral measuring the flux involves
  an inward pointing normal ({\em i.e.}, into the $5d$ bulk).
  }

The degree to which the baryon density is localized depends on how
close the string endpoint is to the boundary of the $5d$ geometry.
The farther the endpoint is away from the boundary,
the more the field lines of $\mathcal A_M$ can spread out,
and hence the more delocalized will be the induced image current
$J_{\rm baryon}^{\mu}$.
In the limit where the radial coordinate $\u$ of the string endpoint
is far from the horizon,
$\u \ll u_h$, the baryon density will be
localized with a length scale $\sim \u$ \cite{Chesler:2008wd}.
We note that the appearance of the length scale $\u$ in the baryon density
is natural since,
for $\u \ll u_h$, it takes light an amount of time $\sim \u$
to reach the boundary.

If at time $\to$ the string's endpoint is at radial coordinate $\uo \ll u_h$, then
$\bar {\mathbf x} (\to)$ approximately coincides with the spatial position of
the string endpoint ${\mathbf x}_{\rm string} (\to)$  \cite{Chesler:2008wd}.  The string endpoint
can only travel a finite distance before falling into the black hole.
The final spatial coordinate of the string endpoint ${\mathbf x}_{\rm string}(\infty)$ will
exactly coincide with $\bar {\mathbf x} (\infty)$  \cite{Chesler:2008wd}.
We therefore have
\begin{equation}
    \Delta x
    \approx
    |{\mathbf x}_{\rm string}(\infty) -{\mathbf x}_{\rm string}(\to)| \,.
\end{equation}

To make the quantity $\Deltax(E)$ meaningful,
we also need to measure the
quark's energy at time $\to$.  After all, we want
to know how far a quark with a given initial energy can travel.
If the quark has been moving for some time
prior to $\to$, it will have deposited energy into the plasma ---
we must disentangle the
energy deposited in the plasma from the remaining energy of the quark itself.
In the limit where the quark has an arbitrarily large energy which
is localized in a arbitrarily small region of space, separating
the quark's energy from the energy transfered to the plasma
will be unambiguous.

Via Einstein's equations, the presence of the string will also perturb the $5d$
geometry.  As in the electromagnetic problem, the perturbation in the geometry
will induce a corresponding perturbation in the $4d$ stress tensor on the boundary
\cite{Brown:1992br, Skenderis:2000in}.
The string itself has a conserved energy.  For the states we
consider in this paper, the endpoints of the string are very close to the boundary
at time $\to$
and hence have a very high gravitational potential energy.
In Section \ref{asymsol} we argue
that the energy contained near a string endpoint
scales like $1/u^3_*$.  It is this \textit{UV sensitive}
part of the string energy that we identify with the energy of a quark.
Via the gravitational bulk-to-boundary problem
[also illustrated in the cartoon of Fig.~(\ref{bulk2boundary})]
the high energy density near the string endpoint gets mapped onto
a region of $4d$ space which coincides with the location of the
quark's baryon density.
Therefore, at time $\to$, we only need to compute the
part of the string's energy which diverges in the
$\uo \rightarrow 0$ limit in order to
identify the energy of the corresponding quark.

\section{Falling strings}
\label{fallingstrings}

The dynamics of a classical string are governed by the Nambu-Goto
action
\begin{equation}
\label{ng}
S_{\rm NG} = -T_0 \int d\tau \> d\sigma \sqrt{-\gamma} \,,
\end{equation}
where $T_0 = \sqrt{\lambda}/(2 \pi L^2)$ is the string tension,
$\sigma$ and $\tau$ are worldsheet coordinates,
and $\gamma \equiv \det \gamma_{ab}$ with $\gamma_{ab}$ the
induced worldsheet metric.
The string profile is determined by a set of
embedding functions $X^M(\tau,\sigma)$.   In terms of these functions
\begin{equation}
\gamma_{ab} = \partial_a X \cdot \partial_b X \,,
\end{equation}
and
\begin{equation}
-\gamma = (\dot X \cdot X')^2 - \dot X^2 X'^{\,2} \,,
\end{equation}
where $\dot X^M \equiv \partial_\tau X^M$ and $X'^M \equiv \partial_\sigma X^M$.

The equations of motion for the embedding functions,
as well as the requisite open string boundary conditions,
follow from demanding vanishing variation of the Nambu-Goto action.
Explicit forms of the resulting equations of motion,
for the class of configurations we will consider, are shown
below in Section \ref{asymsol}\,.
The boundary conditions for the open string
require that its endpoints move at the local speed of
light and that their motion is transverse to the string.

We will limit attention to configurations for which
the string embedding only has non-zero components along a single Minkowski
spatial direction which we will denote as $\hat x$.
We also restrict attention to
initial conditions such that at worldsheet time $\tau = 0$,
the string is mapped into a single point in spacetime.
Explicitly,
\begin{align}
t(0,\sigma) = t_{\rm c}\,,
\ \
x(0,\sigma) = x_{\rm c}\,,
\ \
u(0,\sigma) = u_{\rm c}\,,
\end{align}
where the numbers $t_{\rm c}$, $x_{\rm c}$, and $u_{\rm c}$
specify the $5d$ spacetime location of the string creation event.
The remaining initial data are the velocity profiles at $\tau = 0$,
namely
$\dot t$, $\dot x$ and
$\dot u$ as functions of $\sigma$.
One of these three velocity
functions may be eliminated via gauge fixing.  For example one
may choose the gauge $\tau = t.$

We will be interested in choosing initial data such that the subsequent
evolution leads to configurations in which the two string endpoints
propagate away from each other and become well separated before
falling into the horizon.
Choosing a frame in which the total spatial momentum of the string vanishes,
this implies that one half of the string will carry a large positive 
spatial momentum in the $\hat x$ direction,
while the other half carries a large negative spatial momentum.
We also require that the velocity profiles are smooth near
the string endpoints.
A sufficient condition is that the Fourier series of the velocity profiles
be rapidly convergent (pointwise).
For brevity, we will refer to string configurations satisfying
these conditions as ``reasonable''.
We postpone more detailed discussion of our specific choices
of velocity profiles used for numerical studies to Section~\ref{numerics}\,.

\begin{figure*}[t]
\includegraphics[scale=1.5]{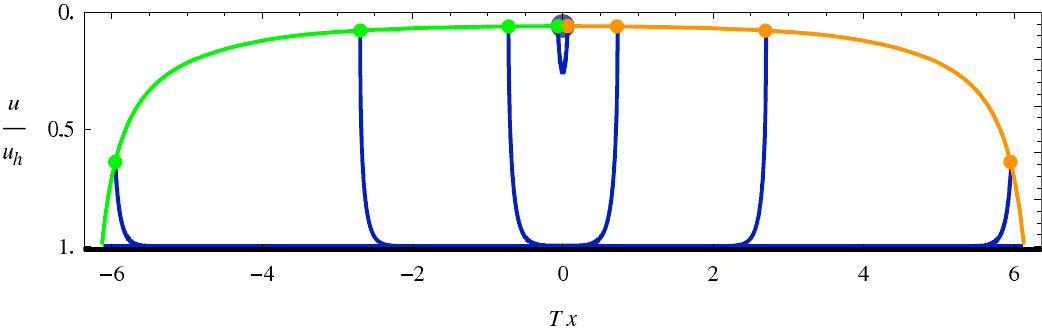}
\caption
  {
  \label{finiteTsymSeq}
  A typical falling string studied in this paper,
  plotted in blue at four different instants in time.
  The string is created at a point and, as time passes,
  evolves into an increasingly extended object.
  Well after the creation event,
  but long before the plunge into the horizon,
  the string profile approaches a universal \textit{null string}
  configuration which is largely insensitive to the initial conditions.
  Consequently, the string endpoint trajectories,
  shown in green and yellow,
  approach null geodesics.
  }
\end{figure*}

\subsection{Approximate solutions}
\label{asymsol}

Consider a string created in the distant past.
In particular, take the radial coordinate of the creation event
to be arbitrarily small,
$u_{\rm c} \rightarrow 0$.
As time progresses, the string evolves from a point into an extended object and the
endpoints of the string fall toward the horizon.
An example of such string evolution (numerically computed) is
shown in Fig.~\ref{finiteTsymSeq}\,.
As we discuss in detail below, in the limit
$u_{\rm c} \rightarrow 0$, the string endpoints can be made to
travel arbitrarily far in the spatial $\hat x$ direction before
falling into the black hole.
Our strategy in this subsection is to construct an approximate
solution to the string equations of motion which will provide
a good description for times sufficiently long after the initial creation
event but well before the string endpoints reach the horizon.
This will be possible because, as we will discuss, 
at times well after the creation event
but long before the final ``plunge'', typical string configurations
approach near-universal forms which are characterized by only
a few parameters.
This observation will allow us to prepare states illustrating
universal features and understand the resulting physics of quark energy loss,
without requiring a detailed
description of the early-time dynamics responsible for the
production of the quark-antiquark pair.

For reasonable falling string solutions,
we will see that the endpoint motion is well-approximated
by the trajectory of a light-like geodesic.
Equations for null geodesics in the AdS-BH geometry are easy work out.
For motion in the $x$-$u$ plane, one finds
\begin{subequations}
\begin{eqnarray}
\label{dxgeo}
\left (\frac{dx_{\rm geo}}{dt} \right )^2 &=& \frac{f^2}{\r^2}\,,
\\ \label{dugeo}
\left(\frac{du_{\rm geo}}{dt} \right )^2 &=& \frac{f^2 \left(\r^2 - f\right)}{\r^2} \,,
\end{eqnarray}
\label{dgeo}%
\end{subequations}
where $\r$ is a constant which determines
the initial inclination of the geodesic in the $x$-$u$ plane
and, more fundamentally, specifies
the conserved spatial momentum associated with the geodesic,
$f(u)^{-1} dx_{\rm geo}/dt = \r^{-1}$.
Moreover, we have
\begin{equation}
\label{dxgeodu}
\left (\frac{dx_{\rm geo}}{du} \right )^2 = \frac{1}{\r^2 - f} \,.
\end{equation}
{}From this equation, one sees that geodesics 
which start close to the boundary, at $u = u_* \rightarrow 0$,
can travel very far in the $\hat x$ direction
provided $\r^2 \approx f(u_*) \rightarrow 1$.  
In particular,
the total spatial distance such geodesics travel before
falling into the horizon scales like $u_h^2/\uo$.

We will be interested in string configurations where
the spatial velocity of the string endpoint is close to
the local speed of light for an arbitrarily long period of time
(since this will maximize the penetration distance).
Because open string endpoints must always travel at the
speed of light, the velocity in the radial direction must be
small and correspondingly, the radial coordinate of the
string endpoints will be approximately constant
for an arbitrarily long period of time.
As the string endpoints become more and more widely separated,
the string must stretch and expand. 
For reasonable string profiles,
this implies that short wavelength perturbations in the initial structure
of the string will 
stretched to progressively longer wavelengths,
resulting in a smooth string profile at late times.%
\footnote
  {%
  ``Unreasonable'' string profiles can have structure on
  arbitrarily short wavelengths.
  While the initial structure will be inflated as time progresses,
  because the string endpoints can only travel a distance
  of order $u_h^2/u_{\rm c}$ before reaching the horizon,
  one can always cook up initial conditions such that
  fluctuations in the string profile never become small during
  this time interval.
  We will avoid such unreasonable initial conditions in this paper.
  }
Moreover, as the string
endpoints separate, the middle of the string must fall
toward the event horizon.
This occurs on a time scale $\Delta t$ of order $u_h$.
(This scale sets the infall time of a particle released at rest
at the boundary, or of a null geodesic with $\r > 1$.)

The origin of this behavior can also be understood as follows.
Consider the string at some time $t$ shortly after the creation event.
It will have expanded to a size $\sim t$.
By construction,
one half of the string will have a positive large momentum in
the spatial $\hat x$ direction,
while the other half has negative $\hat x$ momentum.
The spatial momentum density must be highly inhomogeneous so that
the two endpoints
move off in opposite directions.
As time progresses, the parts of the string with the
highest momentum density will remain close to a string endpoint.
Portions of the string with low spatial momentum density will lag behind
the endpoints (in terms of motion in the $\hat x$ direction)
and fall relatively unimpeded toward the horizon.
Thereafter,
the outer parts of the string will continue moving in the $\pm \hat x$
direction while the endpoints slowly fall.
This general behavior is clearly seen in Fig.~\ref{finiteTsymSeq}\,.

With the above qualitative picture in mind,
we now turn to the explicit analysis.
To simplify the discussion, focus attention on one half of the string.
We may choose worldsheet coordinates $\tau = t$ and $\sigma = u$,
so that the embedding functions are determined by a single
function $x(t,u)$.
The domain in the $(t,u)$ plane in which $x(t,u)$
is defined is bounded by a curve $\u (t)$
which defines the trajectory of the string endpoint.
The location of this curve
is fixed by the open string boundary conditions.
With our choice of worldsheet coordinates,
these boundary conditions are simply
\begin{subequations}
\label{bothboundaryconditions}
\begin{eqnarray}
\label{lightlike}
G_{MN} \,\frac{d X^M}{d t} \, \frac{d X^N}{d t} &=& 0,
\\ \label{transverseT}
G_{MN} \,\frac{d  X^M}{d t}\, \frac{\partial X^N}{\partial u} &=& 0,
\end{eqnarray}
\end{subequations}
where the total time derivatives denote derivatives evaluated
along the curve $\u(t)$.
As noted earlier,
these conditions just express the constraints that
the speed of the string endpoint equal the local speed of light with
a velocity which is transverse to the string.

With our choice of worldsheet coordinates, the determinant of the 
worldsheet metric is
\begin{equation}
\label{detgamma}
\gamma = \frac{L^4}{u^4 f} \left (f^2 x'^{\,2}-\dot x^2 +f \right ).
\end{equation}
Substituting this into the Nambu-Goto action (\ref{ng}), one finds the following 
equation of motion for the embedding function $x(t,u)$,
\begin{align}
\label{xeqm}
0 ={}& 2 u \left(1{+}f x'^2 \right ) \ddot x
- 2 u f \left(f {-} \dot x^2 \right) x''
- 4 u f x' \dot x \dot x' 
\nonumber \\ & {}
+ 4 f \left [2-f \left (1{-}x'^2 \right ) \right ] x'
-4 (3{-}2 f ) x' \dot x^2 \,.
\end{align}

We want to construct an approximate solution for configurations
where the string endpoint reaches the event horizon
after traveling an arbitrarily large spatial distance.
We imagine first matching our approximate solution onto
an exact string solution at a time $\to$.
At time $\to$, suppose that one
endpoint of the string has fallen to the radial coordinate $\uo$.
Because we are considering $u_{\rm c} \rightarrow 0$,
we can always take
\begin{equation}
    u_{\rm c} \ll \uo \ll u_h \,.
\end{equation}
For reasonable initial conditions,
we can take $\to$ sufficiently large so that the string
will be close to a quasi-stationary configuration in which
the string profile uniformly translates while
the string endpoint slowly falls.
In other words,
we seek a perturbative solution to Eq.~(\ref{xeqm}) of the form
\begin{equation}
\label{perturbativesolution}
x(t,u) = x_{\rm steady}(t,u) + \delta x (t,u) + O\big((\delta x)^2\big)\,,
\end{equation}
where
\begin{equation}
\label{sstate}
x_{\rm steady}(t,u) = \r t + x_0(u)
\end{equation}
is a stationary solution to the equations of motion (\ref{xeqm})
and $\delta x(t,u)$ is a first order perturbation satisfying
\begin{equation}
\label{smallness}
| \delta \dot x(t,u) | \ll \r, \quad | \delta x(t,u) | \ll | x_{0}(u) |,
\end{equation}
for all $t>\to$ and all $u > \u(t)$.

At this point,
the constant $\r$ appearing in the stationary solution
$x_{\rm steady}$ is logically independent from the parameter
$\r$ characterizing null geodesics [{\em c.f.} Eq.~(\ref{dgeo})],
but we will shortly see that the endpoint trajectory of
the stationary solution $x_{\rm steady}$ is in fact
directly related to the null geodesics discussed above.

The endpoint trajectory may similarly be represented as
a zeroth order curve plus a first order correction,
\begin{equation}
\label{us}
\u(t) = \u_0(t) + \delta \u(t) \,,
\end{equation}
where $\u_0(t)$ is the endpoint trajectory when $\delta x(t,u) = 0$.

The function
$\delta x(t,u)$ characterizes the perturbations
in the string which have inflated to long wavelengths.
Our basic strategy is to linearize
the equations of motion and boundary conditions in
both $\delta x(t,u)$ and $\delta \u(t)$.
{}From Eq.~(\ref{xeqm}), the equation of motion for $x_0(u)$
is 
\begin{align}
0 ={} & 2 u f \left (\rr^2 - f \right ) x''_0 + 4 f^2 x'^{\,3}_{0}
\nonumber\\ &
+ 4 \big [ (2-f) f - \rr^2 (3 - 2 f) \big ] x'_0 \,.
\end{align}
The general solution to this equation is given
by functions which satisfy
\begin{equation}
\label{x0p2}
\left (\frac{\partial  x_0}{\partial u} \right )^2 = \frac{ u^4 \left (\rr^2 - f \right )}{u_h^4 f^2 \left (1- C f \right)} \,,
\end{equation}
where $C$ is an integration constant.

Neglecting the perturbations $\delta x$ and $\delta \u$,
the boundary conditions (\ref{bothboundaryconditions})
lead to the two endpoint equations
\begin{subequations}%
\begin{align}
\label{bconstraint1}
\left (\frac{\partial  x_0}{\partial u} \right )^2
&= \frac{\r^2 - f}{f^2}\,,
\\ \label{bconstraint2}
\left( \frac{d \, \u_0}{dt}\right)^2
&= \frac{ f^2 \left( \r^2 - f \right)}{\r^2}\,,
\end{align}
\end{subequations}
where all quantities are evaluated at the string endpoint.
Comparing Eq.~(\ref{bconstraint1}) with Eq.~(\ref{x0p2}), we
see that the two conditions agree provided $C = 1$.
In other words, the boundary condition forces the integration
constant $C$ to equal unity.
Furthermore,
comparing Eq.~(\ref{bconstraint2}) with Eq.~(\ref{dugeo}), we see that
the radial motion of the string endpoint in the stationary solution
coincides with that of a light-like geodesic
(when $\r$ of the stationary solution is identified with $\r$ of the geodesic).
Since the speed of the string endpoint necessarily equals the local
speed of light,
this implies that the zeroth order endpoint trajectory, given by
$u = \u_0(t)$ and $x = \xendpt_0(t) \equiv x_{\rm steady}(t,\u_0(t))$,
is precisely a null geodesic.

With $C = 1$, the (negative root of the) differential equation
(\ref{x0p2}) for $x_0(u)$
becomes
\begin{equation}
\label{x0p2A}
\frac{\partial  x_0}{\partial u} = -\frac{ \sqrt {\rr^2 - f }}{f} \,.
\end{equation}
(Taking the negative square root gives a solution which, for $\r > 0$,
trails the endpoint.)
Substituting the steady state solution $x_{\rm steady}$ into
Eq.~(\ref{detgamma}) and using the above differential equation
for $x_{0}(u)$, reveals that the steady state string solution
is one whose worldsheet metric is everywhere degenerate, $\gamma = 0$.
That is,  $x_{\rm steady}$ represents a null string which is everywhere
expanding at the local speed of light.

In the special case $\r = 1$
(which will be of particular interest below),
Eq.~(\ref{x0p2A}) may be integrated analytically.
One finds
\begin{equation}
\label{nullstring}
x_0(u)  = \frac{u_h}{2} \left [ \tan^{-1}\left ( \frac{u}{u_h} \right )+ \half \log\left ( \frac{u_h - u}{u_h + u} \right ) \right ].
\end{equation}
This is the well known trailing string profile of
Ref.~\cite{Herzog:2006gh}.
Similarly, when $\r = 1$ the boundary condition (\ref{bconstraint2})
may be integrated to find $\u_0(t)$.  The solution
is given implicitly by the equation
\begin{equation}
t -\to = -x_0(\u_0) - \frac{u_h^2}{\u_0}+x_0(\uo) +\frac{u_h^2}{\uo} \,.
\end{equation}
But in much of what follows it will be useful to keep $\r$ arbitrary.

Having found the zeroth order solution, we now turn to the
first order correction which describes perturbations to a
stationary null string.
Linearizing Eq.~(\ref{xeqm}) in $\delta x(t,u)$ yields
the equation of motion
\begin{align}
\label{lineareq}
0 ={}
&\rr^2 \, \delta \ddot x
+ f^2 \left( \rr^2 -f \right ) \delta x''
+ 2 \rr f \sqrt{\rr^2 - f} \, \delta \dot x'
\nonumber\\
&{}+
\frac{4 \rr \left( \rr^2 - 2 f + f^2 \right)}{u \sqrt{\rr^2 - f}} \,
\delta \dot x
- \frac{2 f^2  \left( 1 - 2 \rr^2 + f \right)}{u } \, \delta x' \,.
\end{align}
A general solution to this equation
can be constructed explicitly and has the form
\begin{equation}
\label{perturbationsol}
\delta x(t,u) = u_h \left [\varphi(z(t,u)) + g(u)\, \psi(z(t,u)) \right ] ,
\end{equation}
where $\varphi(z)$ and $\psi(z)$ are arbitrary functions,
$g(u)$ satisfies
\begin{equation}
\label{gdef}
g' = \frac{ u^3_h}{u^4} \, \sqrt{\rr^2 - f} \,,
\end{equation}
and the function $z(u,t)$ is given by%
\footnote
  {%
  The overall factors of $\uh$, $\uh^3/u^4$, and $u_*/\uh^2$
  in Eqs.~(\ref{perturbationsol}), (\ref{gdef}) and (\ref{zdef}) 
  are inserted for dimensional consistency and later convenience.
  }
\begin{align}
\label{zdef}
z(t,u) \equiv &\frac{\uo}{u_h^2} \left [x_{\rm steady}(t,u) - x_{\rm geo}(u) \right ] + z_0\,,
\end{align}
with $z_0$ an arbitrary constant.  For the special case of $\r = 1$, one has
$g(u) = -u_h/u$.  [Note that any constant of integration appearing in $g(u)$ can 
be absorbed into the definition of $\varphi(z)$.]

Using Eqs.~(\ref{dxgeodu}) and (\ref{x0p2A}),
one easily finds that
$z(t,u)$ is constant along null geodesics with the constant of motion $\r$.
Moreover, we may choose the constant $z_0$
such that $z$ vanishes on the endpoint trajectory $\u_0(t)$.

The fact that the perturbative solution to the string equations
of motion contains two arbitrary functions $\varphi(z)$
and $\psi(z)$ is to be expected.  As discussed in the previous section,
the required initial data for the evolution of an initially pointlike
string consists of two arbitrary velocity profiles.
Evidently, the information contained in the initial data
gets mapped via the equations of motion onto the two functions
$\varphi(z)$ and $\psi(z)$.

\begin{figure}
\includegraphics[scale=0.20]{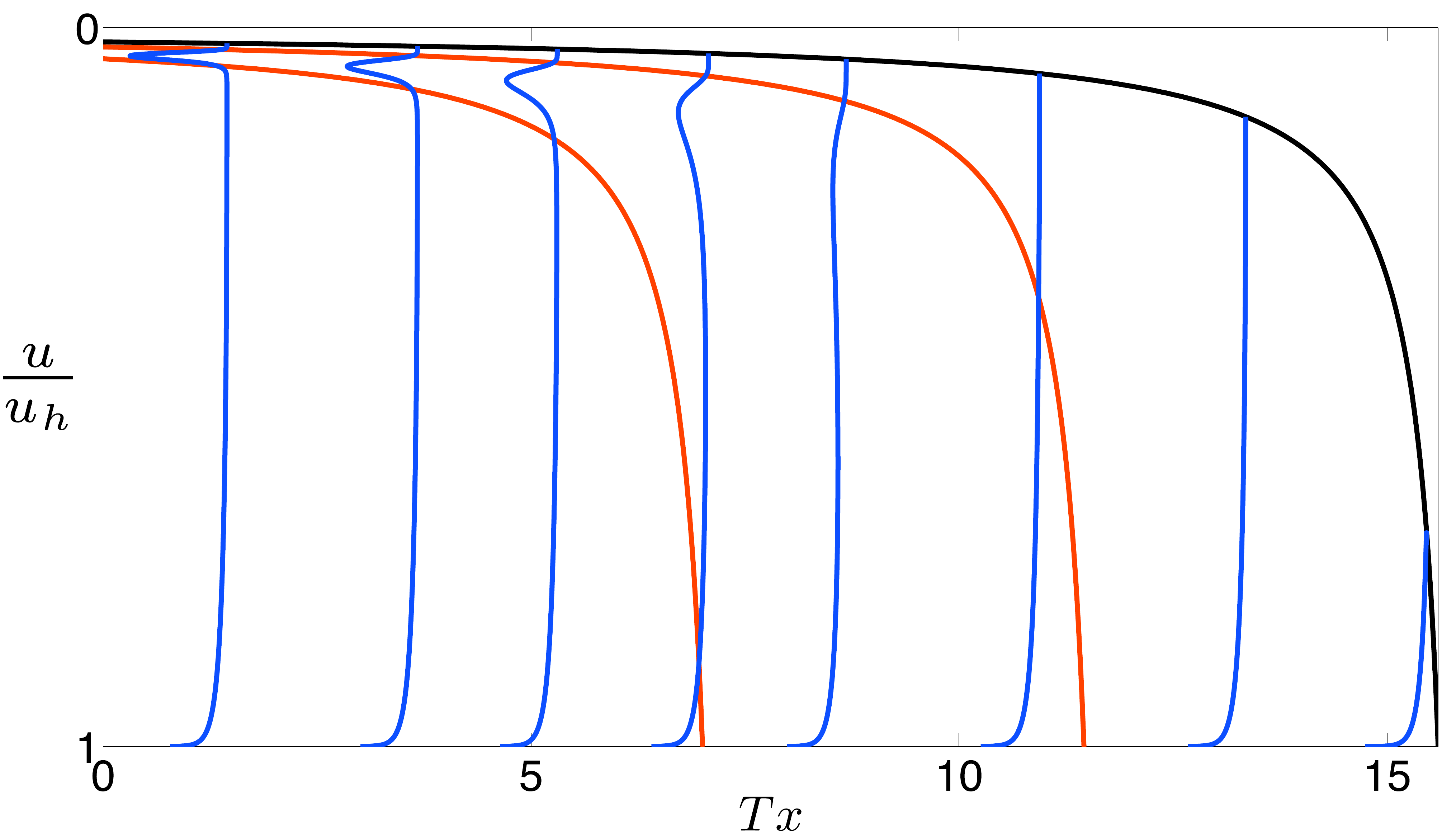}
\caption
  {
  \label{inflation}
  The inflation of a perturbation on a expanding string.
  The thin blue lines show the string at eight different instants of time.
  The uppermost black curve shows the endpoint trajectory.
  The perturbation to the stationary profile is the
  bump initially located close to the string endpoint.
  For clarity, we greatly exaggerate the size of the perturbation.
  The two red curves are lightlike geodesics which enclose the
  perturbation at all times.
  Even though the perturbation is initially highly localized,
  the two geodesics which bound the perturbation
  rapidly separate,
  and correspondingly the size of the perturbation rapidly inflates as it
  falls into the horizon.
}
\end{figure}

It is easy to understand the physical nature of the
perturbations on top of the null string profile.
The null string is everywhere expanding at the local speed of light.
This expansion is analogous to cosmological
inflation --- perturbations defined on top of the null string
at different points are causally disconnected and are
transported along light-like geodesics.
As illustrated in Fig.~(\ref{inflation}), neighboring
geodesics increasingly deviate from each other.
Therefore, as time progresses, the perturbations defined
on top of the null string inflate to long wavelengths.

To finish the first order analysis, we need to find the
correction to the endpoint trajectory.
At linear order in $\delta U$ and $\delta x$,
the boundary conditions (\ref{bothboundaryconditions})
yield the constraints
\begin{equation}
\label{chistar}
\psi(0) = 0 \,,
\end{equation}
and
\begin{equation}
\label{u1dot}
\delta \dot \u = \frac{1}{\r \sqrt{\r^2- f}}
\left [ \left (\varphi'_0 + g \psi'_0 \right) f^2 +\frac{2 \, \u^3_0 \, \delta \u}{ u_h^4} \left (3 f - 2 \r \right ) \right ] ,
\end{equation}
where $\varphi'_0 \equiv \varphi'(0)$,
$\psi'_0 \equiv \psi'(0)$,
and $f$ and $g$ are evaluated at $\u_0(t)$.

\subsubsection{Stopping distance}
\label{stoppingdistance}

If the size of the perturbations on top of the
null string solution are small, then it is easy to compute the
stopping distance --- the
total distance $\Delta x$ traveled by the string endpoint after
time $t_*$.   As the endpoint trajectories of null strings are lightlike
geodesics, $\Delta x$ is simply given by the total spatial distance a geodesic travels.  
Eq.~(\ref{dxgeodu}) gives the result for lightlike geodesics,
\begin{equation}
\label{dxdugeo2}
\frac{dx_{\rm geo}}{du} = \pm \frac{1}{\sqrt{\r^2 - f(u)}}\,.
\end{equation}
The $\pm$ sign reflects the fact that light-like geodesics
can both fall toward the horizon and shoot upwards towards the boundary.
Integrating this equation will yield the total stopping distance.

Reality of Eq.~(\ref{dxdugeo2}) implies that $\r^2 \ge f(u_{\rm min})$,
where $u_{\rm min}$ is the minimal radial coordinate achieved 
along the geodesic trajectory.  
We are interested in trajectories for which $u_{\rm min} \ll u_*$, so 
we require $\r^2 \ge f(u_{\rm min}) \rightarrow 1$.  Physically, this 
corresponds to geodesics which only fall in the vicinity of $t_*$ and thereafter. 
This is sensible in the limit where the string creation point $u_c \rightarrow 0$,
since for
reasonable initial conditions any portion of the motion
in which the string endpoint is moving upward toward the boundary
must occur during the initial transients shortly after the
creation event and well before $t_*$.

For geodesics which only fall, Eq.~(\ref{dxdugeo2}) implies that
\begin{equation}
\Delta x = \int^{u_h}_{\uo} \frac{du}{\sqrt{\r^2 - f(u)}} \,.
\label{eq:Deltax}
\end{equation}
With the 
restriction that $\r \ge1$, $\Delta x$ is maximized at $\r = 1$.
To leading order in
$u_*/u_h \ll1$, evaluating the integral (\ref{eq:Deltax}) gives
\begin{equation}
\label{geodistance}
\Delta x_{\rm max} = \frac{u^2_h}{\uo}\,.
\end{equation}

We must now relate the stopping distance
to the quark energy
at time $\to$.  To make this meaningful, we
want to estimate the minimum amount of energy
required for a quark to travel a distance $\Delta x$
before thermalizing.
This requires determining how the string energy scales with~$\uo$.

The canonical momentum densities of the string are given by
\begin{subequations}
\begin{align}
\pi^0_M &= - T_0 \, \frac{G_{MN}}{\sqrt{-\gamma}}
\left [ (\dot X \cdot X') X'^N - (X'^{\,2})\, \dot X^N \right ],
\\ \label{flux}
\pi^1_M &= - T_0 \, \frac{G_{MN}}{\sqrt{-\gamma}}
\left [ (\dot X \cdot X') \dot X^N - (\dot X^2)\, X'^N \right ],
\end{align}
\end{subequations}
where
$-\gamma = (\dot X \cdot X')^2 - \dot X^2 X'^{\,2}$.
The energy of the string at time $\to$ is then given by
\begin{equation}
\label{energyA}
E_* = - \int^{u_h}_{\uo} du \> \pi^0_t\,.
\end{equation}

The zeroth-order approximation to the
string solution is a null string, for which $\gamma$ vanishes.
Hence, to describe a finite energy configuration,
it is essential to include the
perturbations to the null string profile.
The determinant of the world-sheet metric will necessarily be proportional to
the size of these perturbations.
At linear order, and for $\r = 1$, we have%
\footnote
  {%
  To linear order in the size of perturbations
  away from the null string, the function $\varphi(z)$
  appearing in Eq.~(\ref{perturbationsol}) does not enter into the determinant 
  of the worldsheet metric.  In other words, perturbations in the steady
  state profile $x_{\rm steady}$ induced by $\varphi(z)$ alone
  preserve (to first order)
  the everywhere null character of the string worldsheet.
  }
\begin{equation}
\label{gammanull}
\gamma = \frac{2 L^4  }{u^4} \,  \psi(z)\,.
\end{equation}
For a timelike worldsheet $\gamma$ must be negative, and hence
we must have $\psi(z) < 0$.
Evaluating the string energy (\ref{energyA})
to linear order in perturbations, one finds
\begin{equation}
\label{energy}
    E_* = \frac{\sqrt{\lambda}}{2 \pi}
    \int^{u_h}_{\uo}
    \frac{ du}{ u^2 f} \, \left[{-2 \psi(z(\to,u))}\right]^{-1/2} \,.
\end{equation}
This expression contains an infrared divergence near the horizon.
This divergence reflects the unboundedly large amount of energy
transfered to the plasma from the quark before
time $\to$.%
\footnote
    {%
    More precisely, the upper limit of the integrals
    (\ref{energyA}) and (\ref{energy}) should not be $u_h$,
    it should be the maximal radial coordinate of any point on the
    string at time $\to$ --- which rapidly approaches $u_h$.
    The contribution to the energy from the region $u \gg u_*$ reflects
    energy transferred to the plasma at times $t \ll \to$.
    }
To extract a meaningful energy which can be associated with the
quark at time $\to$,
we focus on
the UV sensitive part of the integral (\ref{energy}).
Neglecting the IR region is tantamount to cutting off
the radial integral at a radial coordinate $u_{\rm IR} < u_h$.
The UV sensitive part of the string energy is the leftover part of the
integral that diverges as
$\uo \rightarrow 0$.
This is the portion of the string energy that should be identified
with the energy of a localized quark jet at time $\to$.

To minimize the energy (\ref{energy}) for a given value of $\uo$,
one wants to maximize the magnitude of the function $\psi(z)$
which characterizes the fluctuation profile.
However, it is necessary to ensure that the perturbative treatment
remains valid near the string endpoint --- one cannot arbitrarily 
crank up the size of $\psi(z)$ as one must ensure that
the relations (\ref{smallness}) are satisfied.
More physically, we demand a well behaved solution, which is approximately
a steady state solution, in the $\uo \rightarrow 0$ limit.
We remind the reader that, as discussed above in Section~\ref{asymsol}\,,
solutions which are approximately steady state solutions are dual
to long lived quarks.
 
At time $\to$,
the UV sensitive part of the string energy comes from contributions near
the string endpoint.  Hence, we focus our attention on the region in which
$u = w \uo$ with $w = \mathcal O(1)$.  Within this region we have
\begin{equation}
x_0(w \uo) = -\frac{w^3 u_*^3}{3 u_h^2} + \mathcal O(u_*^7).
\end{equation}
Then from (\ref{perturbationsol}) we see that $\delta x$ will scale with the same power of $u_*$ if
\begin{subequations}
\label{scalingrelations}
\begin{align}
    \phi(z) &= \left (\frac{\uo}{u_h}\right )^{3} \tilde \phi(z) \,,
\\ 
    \psi(z)  &=  \left (\frac{\uo}{u_h}\right )^{4} \tilde \psi(z) \,,
\end{align}
\end{subequations}
with $\tilde\phi(z)$ and $\tilde\psi(z)$ functions which remain
bounded as $\uo\rightarrow 0$.  With these scalings, a small
$\delta x$ relative to $x_0$ can always be obtained
by adjusting the overall normalization of $\delta x$ with a
numerical factor which is independent of $\uo$.  Similar
conclusions can also be reached regarding the scaling of $\delta \u$ relative to
that of $\u$.%
\footnote
  {%
  From the differential equation (\ref{u1dot}),
  the scalings of $\varphi$ and $\psi$ imply $\delta \u$ scales 
  like $\uo$.  Therefore, in the $\uo \rightarrow 0$ limit
  the smallness of $\delta \u$ relative to $\u$, which is at most $\uo$, 
  can always be achieved by adjusting the 
  size of $\delta \u$ with a constant independent of 
  $\uo$.
  }

With the above scaling of $\psi$, we see that the UV sensitive part of the 
energy (\ref{energy}) can be written
\begin{equation}
\label{energy2}
E_* = \frac{ u^2_h \sqrt{\lambda}}{ \pi^4 u^3_* } \, \frac{1}{\C^3} \,,
\end{equation}
where
\begin{equation}
\label{Cint}
\frac{1}{\C^3} \equiv \frac{\pi^3}{2} \int_1^{w_{\rm IR}} \frac{dw}{w^2}
\left[{-2 \tilde \psi(z(\to,\uo w))}\right]^{-1/2}\,,
\end{equation}
and $w_{\rm IR} = u_{\rm IR}/\uo$.  
By the scaling relations (\ref{scalingrelations}),
the constant $\C$ is finite and independent of $\uo$
in the $\uo \rightarrow 0$ limit.

After using the result (\ref{energy2}) to express $u_*$ in terms of $E_*$,
Eq.~(\ref{geodistance}) yields
\begin{equation}
\label{Deltax}
\Delta x_{\rm max}(E_*) = \frac{\C}{T} \left ( \frac{E_*}{T \sqrt{\lambda}} \right )^{1/3}\,.
\end{equation}
We reiterate that the
$E^{1/3}_*$ scaling is the \textit{maximum} possible power of energy
consistent with the perturbative solution we have derived.  In particular,
it is the maximum power consistent with a string profile which is approximately
a steady state profile.

\subsection{Numerical string solutions}
\label{numerics}

It is instructive to complement the above analytic analysis
with explicit examination of numerically computed string solutions.
We wish to verify explicitly that
(\emph{i}) strings whose endpoints travel far in the Minkowski spatial
are well approximated by null strings,
(\emph{ii}) the endpoint trajectories of such strings are well
approximated by light-like geodesics with $\r = 1$,
and (\emph{iii}) the maximum distance $\Delta x$ that a string
endpoint can travel scales like $E^{1/{3}}$.
The first two points have already been demonstrated numerically in
Ref.~\cite{Chesler:2008wd}, but we have extended that earlier analysis
by exploring a larger sample of initial conditions.
This larger sample size is what allows us to address point (\emph{iii}).
As we are using the same numerical methods as in Ref.~\cite{Chesler:2008wd},
this section closely parallels the analogous discussion there.

To gain insight into the predicted $E^{1/{3}}$ scaling of $\Delta x$,
we solve the string equations of motion
numerically for a variety of initial conditions, and
plot the penetration depth as a function of energy.
As discussed below, we indeed find that the scaling relation (\ref{Deltax})
represents an upper bound on how far a string endpoint can travel for a given
initial energy.  Moreover, the numerical solutions provide a direct estimate
of the constant $\C$ in the bound (\ref{Deltax}).

For reasons discussed below (and earlier in Ref.~\cite{Herzog:2006gh}),
in our numerical analysis we have found it convenient to use the Polyakov
string action.  The Nambu-Goto action is
classically equivalent to the Polyakov action
\begin{equation}
    S_P=-\frac{T_0}{2}\int d^2\sigma \> \sqrt{-\eta} \, \eta^{ab}\,
    \partial_a X^M\partial_b X^N \, G_{MN}\,,
\end{equation}
where one has introduced additional degrees of freedom in $\eta_{ab}$,
the worldsheet metric.  Varying the Polyakov action with respect
to $\eta_{ab}$ generates the constraint equation
\begin{equation}
\label{metricConstraint}
\gamma_{ab}=\coeff{1}{2}\,\eta_{ab} \, \eta^{cd} \, \gamma_{cd}\,.
\end{equation}
This implies that
\begin{equation}
\label{metricEquiv}
\sqrt{-\gamma} \, \gamma^{ab}=\sqrt{-\eta} \,\eta^{ab}\,,
\end{equation}
so that the worldsheet metric differs from the induced metric only by a
Weyl transformation,
\begin{equation}
\eta_{ab}(\tau,\sigma)= e^{2\omega(\tau,\sigma)} \, \gamma_{ab}(\tau,\sigma)\,.
\end{equation}
When Eq.~(\ref{metricEquiv}) is substituted back into the
Polyakov action, one recovers the Nambu-Goto action.

The equations of motion for the embedding functions $X^M$ as well as
the open string boundary
conditions follow from variation of the
Polyakov action with respect to
the $X^M$.  Specifically, one finds
\begin{align}
\label{stringEoM}
    &\partial_{a}
    \big[ \sqrt{-\eta}\,\eta^{ab} \, G_{MN}\, \partial_{b}X^{N}  \big]
\nonumber\\&\quad{}
    =
    \half\sqrt{-\eta}\,\eta^{ab}
    \frac{\partial G_{NP}}{\partial X^{M}} \,
    \partial_{a}X^{N}\partial_{b}X^{P},
\end{align}
together with the boundary conditions
\begin{equation}
\label{openBC}
\pi^\sigma_{M}(\tau,\sigma^*) =0\,.
\end{equation}
Here $\sigma=\sigma^*$ denotes a string endpoint and
$\pi^{\sigma}_M$ is the canonical momentum flux on the
worldsheet,
\begin{equation}
    \pi^\sigma_{M}(\tau,\sigma)
    \equiv
    \frac{\delta S_{\rm P}}{\delta X'^M(\tau,\sigma)}
    =
     -T_0 \sqrt{-\eta} \, \eta^{\sigma a} \, G_{MN}\partial_a X^N \,.
\end{equation}

We can fix the coordinate parametrization $(\tau,\sigma)$ by
choosing the worldsheet metric $\eta_{ab}$. As in
Refs.~\cite{Herzog:2006gh,Chesler:2008wd}, we have found it
convenient to choose $\eta_{ab}$ to be of the form
\begin{equation}
\label{wsMetric}
\|\eta_{ab}\|=\left(\begin{array}{cc} -\Sigma(x,u) & 0 \\ 0 & \Sigma(x,u)^{-1}
\end{array}\right).
\end{equation}
We refer to $\Sigma$ as the stretching function, which
we take to be a function of $x(\tau,\sigma)$ and $u(\tau,\sigma)$ only.
The choice of the stretching function $\Sigma$ is a choice of gauge.
Changes in $\Sigma$ lead to different embedding functions $X^M(\tau,\sigma)$,
but do not affect the geometry of the target worldsheet.
With a worldsheet metric of the form (\ref{wsMetric}),
the constraint equations Eq.~(\ref{metricConstraint}) read
\begin{subequations}
\begin{align}
\label{con1}
\dot X \cdot X' = 0\,,
\\ \label{con2}
\dot X^2 + \Sigma^2 X'^2 = 0\,.
\end{align}
\end{subequations}
Since we choose to study strings with point-like initial
conditions, the $\sigma$ derivatives $X'^M$ are initially zero.
Hence, we must choose initial time derivatives $\dot{X}^M$
which are consistent with the constraint
(\ref{con2}) and the boundary condition (\ref{openBC}).
We may satisfy the constraint (\ref{con2}) by fixing $\dot{t}$
in terms of $\dot{x}$ and $\dot{u}$ via
\begin{equation}
f\,\dot{t}^{\,2}=\dot{x}^2+\frac{\dot{u}^2}{f}\,.
\end{equation}
To satisfy the open string boundary condition (\ref{openBC})
at worldsheet time $\tau = 0$, we choose $\dot{x}$ and
$\dot{u}$ so that
\begin{equation}
\label{finalBC}
\dot{x}'(0,\sigma^*)=\dot{u}'(0,\sigma^*)=0\,.
\end{equation}
The set of pointlike initial conditions then reduce to the choice of
two functions $\dot{x}$ and $\dot{u}$ obeying
Eq.~(\ref{finalBC}), together with the initial radial coordinate $u_{\rm c}$.

To understand why it is preferable to start from the
Polyakov action instead of the Nambu-Goto action when
solving numerically for the string dynamics,
note that the equations of motion (\ref{stringEoM})
contain relative factors of $(-\eta)^{-1}$ between different terms.
Consequently, the string equations become
singular whenever $\sqrt{-\eta}\rightarrow 0$.
If we choose the worldsheet
metric to be the induced metric,
which is equivalent to starting from the Nambu-Goto action,
then the equations of motion become singular as any part
of the string approaches a lightlike configuration.
This always happens at late times as the string accelerates
toward the black brane.
By using the Polyakov form of the string action, and exploiting
the freedom to choose a worldsheet metric of the form (\ref{wsMetric}),
we may rescale the worldsheet metric so that the equations
of motion remain well-behaved everywhere on the worldsheet.

The energy of the string is a conserved
quantity and can be computed from the data
defining the initial conditions.
With
\begin{equation}
\pi_t^{\tau}(\tau,\sigma) = \frac{\delta S_{\rm P}}{\delta \dot t(\tau,\sigma)}
\end{equation}
denoting the conserved canonical energy density,
the total string energy is given by
\begin{equation}
E_{\rm string} = - \int_0^\pi d \sigma \> \pi^\tau_{t}(0,\sigma)\,.
\end{equation}
Expressing this more explicitly in
terms of the initial data, one finds that
\begin{equation}
E_{\rm string} = \frac{\sqrt{\lambda}}{2 \pi} \,
\frac{f(u_{\rm c})}{\Sigma(x_{\rm c} ,u_{\rm c}) \, u_{\rm c}^2}
\int_0^{\pi} d \sigma \> \dot t(0,\sigma)\,.
\end{equation}

\subsubsection{Initial conditions and numerical results}

We consider a two-parameter family of initial conditions.
Inspired by the strings studied in Ref.~\cite{Chesler:2008wd},
we choose
\begin{subequations}
\begin{eqnarray}
\label{xIC}
\dot{x}(0,\sigma) &=&A \, u_{\rm c} \cos \sigma\,, \\
\label{uIC}
\dot{u}(0,\sigma) &=& u_{\rm c}\sqrt{f(u_{\rm c})} \, (1-\cos 2\sigma)\,,
\end{eqnarray}
\label{IC}%
\end{subequations}
and also take $x_{\rm c}=0$.  As $u_{\rm c} \rightarrow 0$ and 
$A \,u_c \rightarrow \infty$, these initial conditions generate strings 
whose endpoints travel arbitrarily far in the Minkowski spatial 
directions before falling into the black hole.  Moreover, 
since $\dot{x}(0,\sigma)$ is antisymmetric about $\sigma=\pi/2$ while 
$\dot{u}(0,\sigma)$ is symmetric, these strings are symmetric about $x=0$ 
at all times.  These states therefore have zero total spatial momentum,
but each half 
of the string has an energy and momentum that scale linearly with $A$
for large $A$.

As in Refs. \cite{Herzog:2006gh,Chesler:2008wd}, we choose a stretching
function so that gradients of the embedding functions are small at all
times during the string's evolution.
We found by trial and error that stretching functions of the form
\begin{equation}
\Sigma(x,u)=
\left[1+\left(\frac x{\pi T}\right)^{2}\right]^m
\left(\frac{1-u/u_h}{1-u_{\rm c}/u_h}\right)
\left(\frac{u_{\rm c}}{u}\right)^2.
\label{eq:stretch}
\end{equation}
were adequate to generate long-lived strings with a variety of initial 
conditions.  In this work, the free parameter $m$ was usually chosen to be $0.02$.

\begin{figure*}[htc]
\includegraphics[scale=1.5]{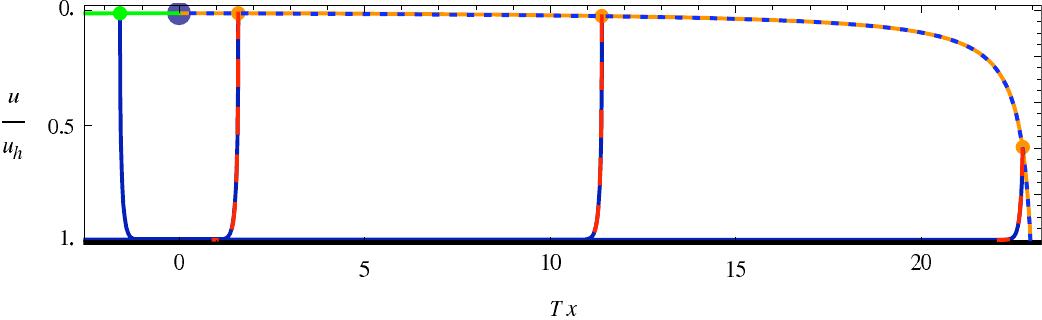}
\caption
  {
  \label{finiteTsymSeq2}
  A plot of a numerically computed string
  at three different times, overlain with the analytic null string
  approximation (\ref{sstate}).
  The string was created at a point at time $t = 0$ with the initial conditions
  (\ref{IC}), for $u_{\rm c} = 0.014 \, u_h$ and $A =2400$.
  The corresponding energy
  $E = E_{\rm string}/2 \simeq 85\,700 \, \sqrt\lambda \,T$.
  The numerical string, shown as the solid blue curve,
  is plotted at successive times $t_1=1.6/ T,$ $t_2=11.4/ T,$ and $t_3=22.8/ T$,
  and the
  corresponding null string, shown as the dashed red curve, is plotted
  at the same times.
  The solid green and orange curves represent the numerically computed
  endpoint trajectories,
  and the overlain dashed blue curve shows
  the geodesic fit to the endpoint trajectory with geodesic parameter $\xi = 1$.
  The null string approximation agrees very well
  with the numeric string configuration at times $t \gtrsim $ a few $u_h$,
  and the null geodesic curve likewise tracks the endpoint trajectory
  very accurately.
}
\end{figure*}

In terms of the initial conditions (\ref{IC}) and
the stretching function (\ref{eq:stretch}), the string's
energy evaluates to
\begin{equation}
E_{\rm string} = \frac{\sqrt{\lambda}}{2\pi}\frac{\sqrt{f(u_{\rm c})}}{u_{\rm c}}
\int_0^{\pi}d\sigma \sqrt{A^2 \cos^2 \sigma+(1-\cos 2\sigma)^2}\,.
\end{equation}
Because the string corresponds to a quark-antiquark pair
in the dual field theory, $E_{\rm string}$ should be regarded as
twice the initial energy of a single quark.  We emphasize that 
our strings are symmetric about $x=0$ so that each half-string is 
approximately dual to a single dressed quark.   In the following we therefore use
\begin{equation}
E \equiv \half \, E_{\rm string} \,,
\end{equation}
when discussing the dynamics of a single string endpoint.

As in Refs. \cite{Herzog:2006gh,Chesler:2008wd},
we used the Mathematica routine \texttt{NDSolve}
to integrate the string equations of motion
(\ref{stringEoM}) numerically.
We chose the parameters $A$ and $u_{\rm c}$ in our initial conditions
(\ref{IC}) from the intervals
\begin{subequations}
\begin{eqnarray}
A &\in& [1000,9000]\,, \\
\frac{u_{\rm c}}{u_h} &\in& [0.00867,0.019]\,.
\end{eqnarray}
\end{subequations}
This generates half-strings with energies in the interval
\begin{equation}
\frac{E}{\sqrt{\lambda}T} \in [26316,519031]\,,
\end{equation}
allowing us to study states with energies much larger than the characteristic 
scale $\sqrt{\lambda}T$
(while simultaneously achieving high numerical precision). 

Our data runs stepped through this parameter space by fixing $u_{\rm c}$
and then generating strings for many values of $A$.
As a result, we were able to generate over a
thousand string worldsheets and measure the associated stopping distances
and energies.
Our data are summarized in Fig.~\ref{allData}\,.
Each distinct line of data points in the plot
comes from a single choice for $u_{\rm c}$.

\subsubsection{Comparison to the approximate string solution}
\label{compare}

Fig.~\ref{finiteTsymSeq2} displays a typical numerically generated string
at three different coordinate times. 
On top of the numerical string profiles,
we also plot the null string (\ref{sstate}) with $\r = 1$.
Also shown in the figure
are the endpoint trajectories overlain with the
corresponding null geodesic with $\r = 1$.
As is evident from
the figure, the null string provides an excellent
approximation to the numerical string profiles for times
$t$ which are a few $u_h$ or larger,
and the difference between the actual endpoint trajectory
and the null geodesic approximation is imperceptible.

To further elucidate the quality of the geodesic approximation to the
endpoint trajectory, we have computed the quantity
\begin{equation}
\Xi(t)\equiv f(\u(t))\left({d\xendpt}/{dt} \right )^{-1} \,,
\end{equation}
where $\xendpt(t)$ is the $\hat x$-coordinate of the string endpoint.
{}From Eq.~(\ref{dxgeo}), one sees that for a geodesic $\Xi$
is constant and equal to $\r$.
Over the course of the trajectory of the
numerical string shown in Fig.~\ref{finiteTsymSeq2}\,,
$\Xi$ equals 1 to within one part in $10^6$, which is 
the limit of our numerical precision.  Therefore, the endpoint 
path for this string is very well approximated by a $\r=1$ geodesic.
We have verified similar results for many different sets of initial conditions
which correspond to long lived quarks.

\subsubsection{Maximum penetration depth}

Our exploration of a wide range of initial conditions
produced the data for penetration depths shown in
Fig.~\ref{allData}\,.
All our data are consistent with the bound
\begin{equation}
\label{numFit}
\Delta x_{\rm max}(E)
=\frac{0.526}{T}\left( \frac{E}{T\sqrt{\lambda}} \right)^{1/{3}}\,,
\end{equation}
which explicitly confirms
the $E^{1/{3}}$ scaling of the penetration depth
derived in Section \ref{asymsol}\,.
More generally, our numerical results clearly confirm the validity
of the asymptotic analysis leading to the approximate string
solution presented in Section~\ref{asymsol}\,.

It is instructive to estimate the maximum value of $\mathcal C$ based
on the perturbative analysis presented in Section~\ref{asymsol}.
Clearly, from Eq.~(\ref{Cint}) one sees that $\mathcal C$ is maximized
when the quantity $\tilde \psi(z)$ is maximized.
However, the validity of the null string approximation presented in Section~\ref{asymsol}
required $|\tilde \psi(z)| \ll 1$. For $\tilde \psi(z) \sim 1$ the integration appearing
in Eq.~(\ref{Cint}) is order one.  Furthermore, the value of $\mathcal C$ only
depends on the cube root of the integral, so $\mathcal C$ is 
rather insensitive to its precise value.  Therefore, in order to get a crude estimate 
on the value of $\mathcal C$ we set the integral appearing in Eq.~(\ref{Cint}) 
equal to one.  We therefore arrive as the estimate
\begin{equation}
\label{estimate}
\mathcal C \sim \frac{\sqrt{2}}{\pi} \approx 0.45.
\end{equation}
This is remarkably close to the numerically determined value of $0.526$

In addition to the sampling of point-like initial conditions
yielding the data shown in Fig.~\ref{allData}\,,
we have also studied more complicated initial conditions describing
strings which are not point-like at $t=0$.
The results obtained for these initial conditions also
demonstrated the $E^{1/3}$ scaling relation of Eq.~(\ref{Deltax}),
but generally yielded a slightly smaller value of 
$\C$.  All of our numerical results are consistent with the 
value for $\C$ determined from the data shown
in Fig.~\ref{allData}\,, namely $\C=0.526$.  
However, we emphasize that this value, extracted from a finite sampling
of initial conditions, is a lower bound on the true value of $\C$.
It is possible that a wider set of initial conditions will
yield a larger value for $\C$, although because of the 
close agreement with the estimate obtained in Eq.~(\ref{estimate}), we doubt that
the true value is significantly greater than $0.526$.

\section{Discussion}
\label{discussion}

\subsection{Energy loss rate}

As Fig.~\ref{allData} makes apparent,
propagating light quarks in strongly-coupled $\Nfour$ plasma
do not have a unique stopping distance for a given energy.
This result should not be surprising.
Knowledge of the total energy (and momentum) of a
quark-antiquark state is far from a complete specification
of the initial state.
The form of the disturbance in the gauge field
(and other $\Nfour$ SYM fields)
will affect the subsequent dynamics.
In the dual description, this additional information is
encoded in the profile of the string connecting
the quark and antiquark.
Nevertheless, there is a rather simple characterization
of the maximum penetration distance of a quark,
scaling with energy as $E^{1/3}$.

An interesting quantity to consider is the
instantaneous energy loss rate of a light quark.
{}From the $\Delta x \sim E^{1/{3}}$ scaling of the penetration depth,
one might expect that for light quarks the rate of energy loss per distance
traveled, $dE/dx$
(which essentially coincides with $dE/dt$ while the excitation is a good
quasiparticle),
would scale like $E^{{2}/{3}}$.
This expectation turns out to be incorrect, as we now discuss.

Let $f^{\mu}_{\rm drag}(t) = {d p^{\mu}}/{dt}$ denote
the four momentum lost by the quark per unit time.
The long distance hydrodynamic perturbation in the SYM stress
tensor $T^{\mu \nu}_{\rm hydro}$
is determined by the hydrodynamic constituent relations together
with the energy-momentum conservation relation \cite{Chesler:2007sv},
\begin{equation}
\label{energycons}
\partial_\mu T^{\mu \nu}_{\rm hydro} = F^\nu \,,
\end{equation}
with
$
    F^\mu(t,\mathbf x)
    =
    - f^{\mu}_{\rm drag}(t) \,
    \delta^{(3)}(\mathbf x {-} \mathbf x_{\rm quark}(t))
$
the force-density (acting on the plasma)
and $\mathbf x_{\rm quark}(t)$ the quark's trajectory.

As long as the quark's baryon density is well-localized in space,
the energy loss rate may be determined by computing the energy flux
through a sphere $S_R$ of radius $R$ which encloses (nearly) all
of the quark's baryon density.
It is this energy flux which enters
in the force density of Eq.~(\ref{energycons}).
As $1/T$ sets the length scale on which
a hydrodynamic description of the stress tensor perturbation
becomes valid \cite{Chesler:2007sv},
it is natural to take $R \sim 1/T$.
The precise value chosen for $R$ is irrelevant --- during times in
which the quark is a well defined quasiparticle,
its baryon density is localized over a scale $\ll 1/T$
while the distance traveled by the localized baryon density distribution
is $\gg 1/T$.

Using the dual gravitational description, one may compute the
energy flux through $S_R$ by solving the gravitational bulk-to-boundary
problem.
Specifically, one solves Einstein's equations for the perturbation
in the $5d$ geometry due to the presence of the string
and then,
by analyzing the near boundary behavior of the metric perturbation
\cite{Skenderis:2000in,Chesler:2007sv,Friess:2006fk},
extracts the change in the
SYM stress tensor
and uses this result to evaluate the energy flux through $S_R$.
This procedure was carried out for heavy quarks moving at constant
velocity in
Refs.~\cite{Chesler:2007an,Chesler:2007sv,Gubser:2007xz,Gubser:2007ga}.
Carrying out the corresponding analysis for our non-stationary
light quark worldsheets is computationally demanding and
will be left for future work.
However, there is a simple way of extracting the energy loss rate
from the string profile itself.
The energy of a string is a conserved quantity.  The high energy density
near the string endpoint should, as discussed in Section~\ref{lightquark}\,,
be regarded as the energy of the quark.
This energy is transported down the string by an energy
flux $\pi_t^1$ [{\em cf.} Eq.~(\ref{flux})]
towards the event horizon.
This energy flux corresponds to energy transferred from
the quark to the plasma.
Via the holographic bulk-to-boundary mapping,
this conserved flux is mapped onto the energy flux through $S_R$ in the
dual field theory.
Without solving the bulk-to-boundary problem explicitly,
one does not know a-priori 
precisely how to relate the bulk position $(u,t)$ at which one evaluates the
energy flux down the string to a corresponding time and radius $R$
of an energy flux measurement in the field theory.
However, as long as the quark energy loss rate is changing
sufficiently slowly, retardation effects in the gravitational bulk-to-boundary
problem can be neglected and the energy flux through
$S_R$ should be well approximated by evaluating the energy flux down the string
at a spatial distance $\sim R$ from the string endpoint.

\begin{figure}
\includegraphics[scale=0.31]{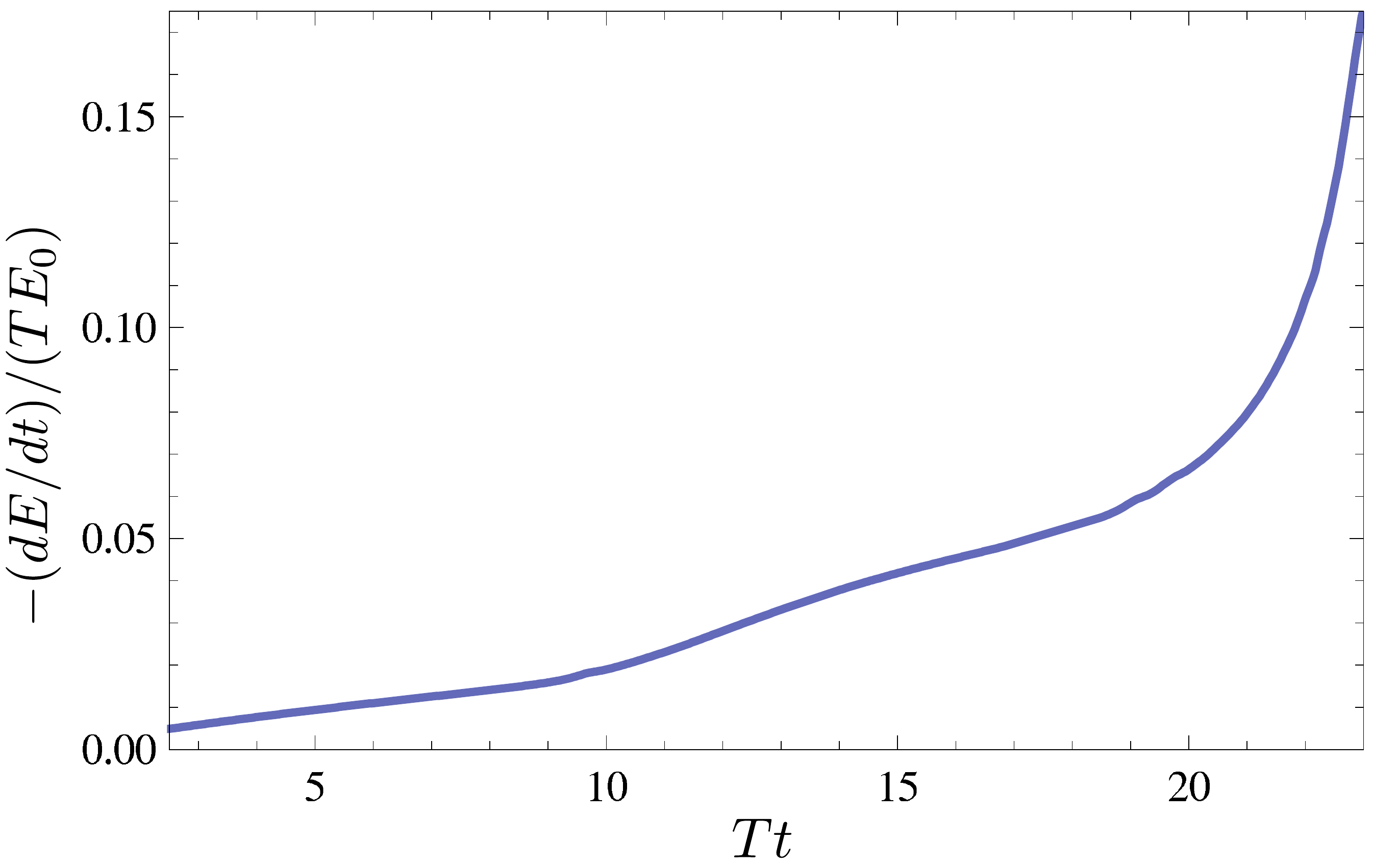}
\caption
  {
  \label{eloss}
  The instantaneous energy loss rate, $dE/dt$,
  of a highly energetic quark,
  normalized by its initial energy $E_0$.
  Instead of decreasing with time,
  as might have been expected,
  the light quark energy loss rate actually increases.
  At times near the thermalization time,
  which for this particular example is $t_{\rm therm} \sim 24/T$,
  the instantaneous energy loss rate grows like
  $dE/dt \sim 1/\sqrt{t_{\rm therm} - t}$.
}
\end{figure}

In Fig.~\ref{eloss}\,, we plot the energy flux flowing down the
numerically generated
string shown in Fig.~\ref{finiteTsymSeq2}\,,
evaluated at a distance of $1.75 /(\pi T)$ from the string endpoint.
For this particular string, the string
endpoint approaches the event horizon at a time $t \sim 24/T$,
which should be regarded as the
thermalization time $t_{\rm therm}$ of the light quark.
As the string endpoint approaches the event horizon,
the baryon density induced on the boundary rapidly
spreads out and diffuses \cite{Chesler:2008wd}.
In the gravitational description,
this is due to the strong gravitational redshift incurred
on the electric field sourced by the string endpoints as they approach
the horizon.%
\footnote
  {%
  More precisely, as the string endpoints approach the horizon,
  the strong gravitational field of the black hole pulls the electric field lines, which are
  sourced by the string endpoints,
  towards the horizon.  This results in the spreading out 
  of the electric field lines and hence a spreading out
  of the induced baryon density on the boundary.
  } 
As is evident from Fig.~\ref{eloss}\,,
the energy flux down the string does not decrease
in a power-law fashion as a naive $E^{{2}/{3}}$
scaling of $dE/dt$ would suggest,
but rather increases monotonically until the thermalization time!

We stress that the precise form of the energy flux down the string
is sensitive to the initial conditions used to create the string.
This is easy to understand from the approximate
analytic string solutions discussed in Section~\ref{asymsol}\,.
These approximate solutions,
which correspond to long-lived quarks, are perturbations of null strings.
The energy flux diverges for a null string.
The finite flux of the complete solution
is determined by the function $\psi(z(t,u))$
[defined in Eq.~(\ref{perturbationsol})]
which characterizes the perturbation $\delta x(t,u)$ on the null string.
This function is not universal, and depends on the initial conditions
used to create the string.

However, the late time behavior of the
instantaneous energy flux is universal.
As is evident from Fig.~\ref{eloss}\,,
near the thermalization time the energy flux down the string
dramatically increases.
This may be understood from our approximate
string solutions.
The energy flux down the string scales like $(-\gamma)^{-1/2}$ where
$\gamma$ is the determinant of the worldsheet metric.
For strings which are small perturbations of null strings,
Eq.~(\ref{gammanull}) shows
$\gamma$ is proportional to the function $\psi(z(t,u))$
characterizing the perturbations.
Near the thermalization time $t_{\rm therm} = u_h^2/\uo$ and at a radial coordinate
$u$ corresponding to a fixed distance $\sim 1/T$ from the string
endpoint, the function $z(t,u)$ behaves like
\begin{equation}
\label{zasym}
z(t,u) = \frac{t - t_{\rm therm}}{\uo} + \mathcal O(\uo/u_h),
\end{equation}
and hence becomes very small as $t \rightarrow t_{\rm therm}$.
By the open string boundary conditions (\ref{bothboundaryconditions})
and (\ref{chistar}),
the function $\psi(z)$ must vanish at the string
endpoint which, as discussed in Section~\ref{asymsol} corresponds to
$z = 0$.  Finiteness of the string energy (\ref{energy})
requires that $\psi'(0)$ be non-zero.
Consequently, near the endpoint one may approximate
$\psi(z) \approx \psi'(0) \, z$.
Neglecting the $\mathcal O(u_*/u_h)$ corrections
in Eq.~(\ref{zasym}), one finds
\begin{equation}
\label{fluxscaling}
\pi_t^1 \sim \frac{1}{\sqrt{t_{\rm therm} - t}}\,.
\end{equation}
We have numerically confirmed the above scaling in the data
shown in Fig.~\ref{eloss}\,.

The late-time behavior (\ref{fluxscaling}) implies that
after traveling substantial distances through the plasma,
the thermalization of light quarks
ends with an ``explosive'' transfer of energy to the plasma.
This behavior is qualitatively similar to the 
energy loss rate of a fast charged particle moving through 
ordinary matter, where the energy loss rate has a pronounced peak 
(known as a ``Bragg peak'') near the stopping point.
This peak in the energy loss rate has its origin in
the energy dependence of cross sections,
which increase with decreasing energy due to the 
conformal nature of Coulomb interactions.

In the gravitational description,
the scaling (\ref{fluxscaling})
becomes valid when the string endpoint starts to fall toward the horizon
({\em i.e.}, when $d\u/d\xendpt$ ceases to be small compared to one).
As shown in Fig.~\ref{finiteTsymSeq2}\,,
this happens relatively abruptly,
so we expect the creation of large amounts
of gravitational radiation to propagate to the boundary and induce a large perturbation in the
SYM stress tensor corresponding to this final burst of energy.
However, we emphasize that because the energy flux flowing down the string
is changing rapidly at late times,
retardation effects in the gravitational bulk-to-boundary problem cannot
be neglected, implying that the result (\ref{fluxscaling})
for the energy flux down the string cannot be directly equated
with the field theory energy flux through a sphere $S_R$.
It would, of course, be interesting to
compute directly the energy flux in the plasma
produced by the light quark jet as it thermalizes.
Evaluation of the required bulk-to-boundary problem
is currently in progress.

It is interesting to speculate on the implications of our results
for heavy ion collisions.
Hard partons produced in the early stages of 
heavy ion collisions can traverse the resulting fireball and deposit energy
and momentum into the medium.
If the partons are moving supersonically,
their hydrodynamic wake will contain a Mach cone whose propagation can
influence the distribution of particles associated with a jet.
If the hard parton under consideration is a very massive quark with mass $m$,
the results of Ref.~\cite{Herzog:2006gh} 
predict an energy loss rate of the form $dE/dx = dp/dt = - \mu p$ where,
for strongly coupled SYM,
$\mu = \frac \pi2 \, \sqrt\lambda \, T^2/m$.
Therefore, the energy loss rate falls exponentially with time ---
heavy quarks in strongly coupled SYM lose the bulk of their energy
in the early portions of their trajectories.
The resulting sound waves, whose amplitudes will be largest at early times,
may traverse much of the fireball before freezeout occurs.
Consequently, sound waves
produced by heavy quarks may be quite sensitive to medium effects and 
may experience substantial attenuation before freezeout.

At least for strongly coupled SYM,
the situation for light quarks is qualitatively different.
As we have demonstrated in Fig.~\ref{eloss}\,,
light quarks lose the bulk of their energy in the 
latter stages of their trajectories.
The resulting hydrodynamic wake will therefore have less time to attenuate
and diffuse than is the case for heavy quarks.
Moreover, because the light quark energy loss rate increases with time,
we expect the amplitude of the corresponding wake to also increase with time.
Because of this, we expect the spectrum and distribution of particles produced
by light quark jets to be qualitatively different
from the behavior of heavy quark jets.

\subsection{Fluctuations}

Throughout our analysis, we have
treated the string dynamics classically.
This approximation is valid in the limit of large 't Hooft coupling $\lambda$.
More precisely, a classical treatment is valid in the limit that
$\lambda \rightarrow \infty$ with $E/\sqrt\lambda T$ finite and fixed.
(Recall that the string energy automatically scales like $\sqrt{\lambda}$.)
However, one would also like to understand when
the classical analysis can be trusted if $\lambda$ is large but fixed.
To determine this, one should compute the size of
quantum fluctuations around the classical string profile and
compare the size of the fluctuations to the classical result.
Natural specific quantities to consider are the fluctuations $\Delta p$
in the quark momentum $\mathbf p$.
These fluctuations are defined by the variances
\begin{equation}
(\Delta p_i(t))^2 = \langle p_i(t)^2 \rangle - \langle p_i(t) \rangle^2 \,.
\end{equation}
If $\Delta \mathbf p$ is not small compared to $\mathbf p$,
then the reliability of the classical calculation is questionable.

Formally, the mean momentum $\mathbf p$ and the connected correlator defining
$(\Delta p)^2$ are both $O(\sqrt{\lambda})$.
Consequently $|\Delta p/p| = O(\lambda^{-1/{4}})$
and vanishes as $\lambda \rightarrow \infty$.
However, for large but fixed $\lambda$
the energy (and time) dependence of $|\Delta p/p|$ can be important.
To see this, consider the case of fluctuations in the momentum of a
heavy quark.
Mean square fluctuations in the longitudinal and transverse components
of the quark's momentum 
grow with time $t$ as
\cite{CasalderreySolana:2007qw,Gubser:2006nz}
\begin{subequations}
\label{eq:fluctuations}%
\begin{eqnarray}
\label{dpl}
 (\Delta p_{\rm L})^2  &\sim&  \sqrt{\lambda}
 \, \gamma^{5/2} \,
 T^3 t\,, \\
\noalign{\hbox{and}}
 \label{dpt}
 (\Delta p_{\rm T})^2  &\sim& \sqrt{\lambda}
 \, \gamma^{1/2} \,
 T^3 t \,,
\end{eqnarray}
\end{subequations}
respectively, where $\gamma \equiv 1/\sqrt{1{-}v^2}$ and
$v$ is the heavy quark velocity.
Therefore, with large but fixed $\lambda$,
the relative size of fluctuations, $\Delta p/p$,
becomes arbitrarily large both at sufficiently late times
and (for longitudinal fluctuations) in the ultra-relativistic limit.

To estimate the size of quantum fluctuations in the light quark's
momentum, we will use the above results for heavy quarks as a rough guide.
This is not unreasonable as
the trailing string profile 
used to compute the above momentum fluctuations coincides,
in the $v \rightarrow 1$ limit,
with the ($\r = 1$) null string derived in Section~\ref{asymsol}\,.
However, care must be taken --- quantum fluctuations on top of the null string,
which is a degenerate solution to the classical equations of motion, diverge.
This is immediately apparent in the formulas (\ref{eq:fluctuations})
which blow up as $v \rightarrow 1$.
To estimate the momentum fluctuations for light quarks using
these results,
we must be able to associate the heavy quark
velocity $v$ (which is always less than 1)
with the size of the \textit{classical} perturbations
$\delta x$ [defined in Eq.~(\ref{perturbativesolution})]
to the null string profile.
To do so, we simply note that the 
trailing string profile for a heavy quark with velocity $v$ coincides
with the null string profile (which is the $v \rightarrow 1$ limit)
up to $\mathcal O(1{-}v^2)$ corrections.
Therefore, for the purpose of a rough estimate,
we identify $1{-}v^2$ with the size of classical
perturbations on top of the null string.
As discussed in Section~\ref{stoppingdistance}\,, if at time
$\to$ the radial coordinate of the string is $\uo$, then the largest%
\footnote
  {%
   Strictly speaking, in Section~\ref{stoppingdistance} we argued that
   $\delta x$ can be no larger than $u^3_*/u_h^3$ only in the
   vicinity of the string endpoint.  However, via the inflationary
   behavior of perturbations defined on top of the null string (as shown
   in Fig.~\ref{inflation}),
   when $t - \to = \mathcal O(1/T)$
   the perturbation in the string
   profile will be determined by the near endpoint perturbations
   at time $t = \to$.
   }
the perturbations to the null string can be is
$\mathcal O(u^3_*/u_h^3)$.
With the identification $1{-}v^2 \Longleftrightarrow \mathcal O(u^3_*/u_h^3)$,
we have
\begin{equation}
(\Delta p_L)^2
\sim
\left (\frac{u_h}{u_*} \right)^{15/4} \,
\sqrt{\lambda} \, T^3 \, t
\,.
\end{equation}
During the portion of the quark's trajectory when it is a well defined
quasi-particle
its momentum, by construction, is large and scales as
$p \sim \sqrt{\lambda}\,{u_h^2}/{u^3_*}$.
We therefore arrive at the estimate
\begin{equation}
\label{dpbyp}
\frac{\Delta p}{p}
\sim
\left (\frac{\uo}{u_h} \right )^{9/8} \frac{\sqrt { T t}}{\lambda^{1/4}}\,.
\end{equation}
In the limit $\uo/u_h \ll 1$
(\textit{i.e.} the high energy limit) we see that
the size of quantum fluctuations are small relative to the classical prediction
for the momentum.
But for fixed values of $\uo$ and $\lambda$,
the $\sqrt t$ growth of the result (\ref{dpbyp}) suggests there will be a time
when quantum fluctuations become large.
However, for the light quarks discussed in this paper, the quarks
only exist as well defined quasiparticles
up until the thermalization time $t_{\rm therm}$.
As discussed in Section~\ref{asymsol}\,, in terms of $\uo$
the thermalization time is simply $t_{\rm therm} = u_h^2/\uo$.
Evaluated at this time, Eq.~(\ref{dpbyp}) implies that quantum fluctuations
are suppressed by a relative factor of $(\uo/u_h)^{5/8}/\lambda^{1/4}$.
This is reassuring.

\subsection{Pair creation, string fragmentation and finite $\Nc$}

In addition to large $\lambda$, we have also assumed from the outset
that $\Nc$ has been sent to infinity.
This limit is what justifies the neglect of quantum fluctuations in the
background AdS-BH geometry.
At finite $N_c$, 
string fragmentation,
backreaction of the string on both the geometry and the $D7$ brane
embedding, and backreaction of the brane on the geometry also have to
be addressed.

Energetically, our string is unstable to breaking into many tiny pieces.
This corresponds to quark-antiquark pair creation in the dual field theory.
The small string coupling
$g_s \sim 1/N_c$ suppresses the amplitude for a string to break.
Correspondingly, the rate of the string decay process is suppressed
by $1/N_c^2$, and hence the time scale for string
fragmentation is, at large $N_c$, parametrically larger than any of the time
scales considered in this work.

Alternatively, the process of string fragmentation
can also be described from the point of view of the $D7$ brane
worldvolume as the spreading of a narrow flux tube, the original
fundamental string, into more and more widely dispersed flux
on the brane.
In terms of the underlying string theory, the quanta of the worldvolume
gauge field are little pieces of open string, so a uniform flux on the
worldvolume is the same as a coherent cloud of little string pieces.
Thinking of the dynamical instability of our string as a result of
breaking into many pieces, or due to spreading into dispersed flux
on the brane,
are just two different descriptions of one and the same process
which is suppressed at large $N_c$.

Large $N_c$ is also what justifies the neglect of backreaction
of the $D7$ brane on the background geometry, as well as
the backreaction of the string on the $D7$ brane embedding and on the geometry.
Note that, as far as large $N_c$ counting
is concerned, the gravitational action scales as $N_c^2$,
the action for the brane embedding and the
worldvolume gauge field scales as $N_c$,
and the Nambu-Goto action describing the
worldsheet of the string is of order 1.
(In addition the three actions scale with the 't Hooft coupling
as 1, $\lambda$ and $\sqrt{\lambda}$,
but for now it is sufficient to focus on the $N_c$ counting.)
The brane is very heavy compared to the string,
but still has a small tension in Planck units.
Consequently, it is consistent to embed the brane in a
fixed background geometry and then consider a string ending on the brane,
without computing the $\mathcal O(1/\Nc)$ suppressed deformation of the brane
which will be induced by the string.

The issue of backreaction becomes more subtle when one solves
for the linearized response of the metric in response to
the string in order to determine the boundary stress tensor.
The order $\Nc^0$ stress energy of the string generates an order $1/N_c^2$
correction to
the metric (since the $5d$ gravitational constant scales as $1/N_c^2$).
Consequently, when evaluating the variation of the 
on-shell gravitational action, the perturbation in the geometry due to the presence of the
string produces an order one contribution to the expectation value
of the stress tensor.

In addition to the string itself, another potential source for
the stress tensor is the gauge field living on
the brane which is sourced by the string endpoint.
The $O(1)$ charge from the string endpoint gives rise to an
order $1/N_c$ gauge field on the brane [as the gauge coupling on the brane is
$\mathcal O(1/N_c)$]. Combining this with the overall $N_c$ of the brane action
would appear to give another order one source in the bulk,
and hence another order one contribution to the expectation value
of the stress tensor out on the boundary. 

However, it is important to note that the leading $\mathcal A_M$
dependent term in the brane stress tensor is quadratic in the worldvolume
gauge field, so that the order $1/N_c$ gauge field on the brane only
produces a stress-tensor contribution of order $1/N_c$.
This conclusion is altered if
there is an order one background electric field on the worldvolume
(as in the original dragging string solution of
Ref \cite{Herzog:2006gh}). In
this case there are contributions to the brane stress tensor linear in the
gauge field sourced by the string, which consequently give rise to
an order one contribution to the stress tensor.
Similar terms also arise if one studies finite, non-zero mass quarks
where an order one background embedding scalar is turned on.
Both of these contributions would need to be included if
one wanted to generalize the stress-energy wake calculations of
Refs.~\cite{Chesler:2007an,Chesler:2007sv, Gubser:2007xz,
Gubser:2007ga} to finite mass quarks.

\subsection{Weak versus strong coupling}

It is natural to ask how
the $E^{1/{3}}$ scaling of the penetration depth in the strongly
coupled limit compares with the analogous result for
weakly coupled plasmas.
When the 't Hooft coupling $\lambda$
(at scales ranging from the temperature $T$ to the projectile energy $E$)
is small,
the energy loss of a high energy parton moving through the plasma
is dominated by near-collinear bremsstrahlung processes.
The rate for an energetic parton (with energy $E \gg T$)
to radiate a gluon which carries away an $O(1)$ fraction of its energy,
while interacting with a typical gauge field fluctuation in the plasma,
scales as $\lambda^2 \,T/\sqrt {E/T}$
\cite {Baier:1996kr,Baier:1998kq,Zakharov:1996fv,Jeon:2003gi,Arnold:2008zu},
up to factors depending logarithmically on the energy,
which we ignore throughout this discussion.
Therefore, the average distance an energetic parton travels between
emission events is $\Delta x_{\rm rad}(E) \sim \sqrt{E/T}/(\lambda^2 T)$.
The square root dependence on energy is due to LPM suppression,
which is a consequence of multiple scattering during the formation time
of a radiated gluon.

Imagine creating a very energetic quark in a localized wave-packet with
mean momentum $\mathbf p$,
and then measuring, at some later time, the total energy or baryon number
contained in a co-moving sphere of size $R \sim 1/T \gg 1/|\mathbf p|$
surrounding the wave-packet.
The opening angle in near-collinear bremsstrahlung emission is
parametrically small,
$\Delta\theta\sim \sqrt\lambda \, (T/E)^{3/4}$.
Therefore, the direction of the leading parton is almost unchanged
by these bremsstrahlung emissions.
Since the speeds of ultrarelativistic excitations 
differ negligibly from the speed of light, this implies that
all the partons produced by a cascade of near-collinear emissions 
have almost identical velocities.
Consequently, near-collinear bremsstrahlung emissions do not significantly
degrade the energy, or baryon number, contained in the co-moving sphere.
As far as gauge-invariant measurements of energy or momentum are concerned, 
the entire collection of near-collinear partons behaves like a single
collective excitation whose energy and momentum is nearly constant.

This effective ``quasi-particle'' picture remains valid until the
typical energy of the partons produced by the cascade ceases to be
large compared to $T$.
The typical penetration depth will equal the radiation length
$\Delta x_{\rm rad}(E)$ summed over the number of levels of showering
which are required to degrade the typical parton energy from $E$
down to $\approx T$.
Since every emission transfers an $O(1)$ fraction of energy to the
emitted gluon, and every produced parton continues to shower,
the typical energy of partons produced by a cascade with $k$ levels
of showering will be of order $E/c^k$ for some $c \approx 2$.
Therefore, 
the number of showerings required to thermalize an extremely
energetic parton grows 
only logarithmically with energy,
and the total penetration depth differs from the radiation length
for the first emission only by an $O(1)$ factor.
The net result is that
the penetration depth $\Delta x(E)$ in a weakly coupled non-Abelian plasma
behaves as $\sqrt{E/T}/(\lambda^2 T)$ times factors depending
only logarithmically on $E/T$.

Presumably, there is a smooth interpolation from weak to strong coupling
in $\Nfour$ SYM.
At intermediate couplings, the maximum penetration depth may be proportional
to $E^{\nu(\lambda)}$, with an exponent $\nu(\lambda)$ which varys
smoothly
from 1/2 as $\lambda \rightarrow 0$ to 1/3 as $\lambda\rightarrow\infty$.
Alternatively, the correct form might be a sum of two distinct contributions,
$
    T\, \Delta x = A(\lambda) \, (E/T)^{1/2} + B(\lambda) \, (E/T)^{1/3}
$,
with $A(\lambda) = \mathcal O(\lambda^{-2})$ and $B(\lambda) = o(\lambda)$
as $\lambda \rightarrow 0$, and
$B(\lambda) = \mathcal O(\lambda^{-1/6})$ and $A(\lambda) = o(\lambda^{-1/6})$
as $\lambda \rightarrow \infty$.
Subleading corrections to the weak-coupling energy loss rate which
are suppressed by powers of the 't Hooft coupling are not known,
and would be challenging to calculate.
And subleading strong-coupling corrections, suppressed by inverse powers
of $\lambda$, are also unknown.
Consequently, there is no way, at present, to determine a preferred
interpolating form.

\subsection{Relation to other work}

In Ref.~\cite{Gubser:2008as}, where the $E^{1/3}$
scaling was first proposed, various guesses for the analog of
$\C$ were given based on different assumptions.  The authors of this work
were interested in calculating the penetration depth of a 
gluon, whose dual description was conjectured to be a folded string.
The relevant string configuration was assumed to be given by
a portion of the stationary trailing string profile of
Ref.~\cite{Herzog:2006gh},
with the string (at any instant of time) coming up from the horizon,
reaching a sharp hairpin at some radial coordinate $u_*(t)$,
and then retracing the same path back down
to the horizon.

The authors of 
Ref.~\cite{Gubser:2008as} estimated the penetration depth
of a gluon by assuming that
the hairpin in the string falls into the horizon along a lightlike geodesic.
Without solving the string equations of motion, 
the parameters of the geodesic were estimated in terms of $\uo$ and $v$.
By relating the these parameters to the string's energy,
the authors of Ref.~\cite{Gubser:2008as} argued that
the maximum penetration depth should scale like
\begin{equation}
\label{ggprresult}
\Delta x_{GGPR} = \frac{\mathcal C_{GGPR}}{T} \left (\frac{ E_*}{2 T \sqrt{\lambda}} \right )^{1/3}.
\end{equation}
The constant $\mathcal C_{GGPR}$ was estimated to be between $0.35$ and $0.41$.

While we have found that the endpoint trajectories of strings corresponding to 
long lived quarks  do follow lightlike geodesics (to quite high accuracy),
the relationship between the parameters of the geodesic 
and the energy of string are rather different than
that presented in Ref.~\cite{Gubser:2008as}.  
In contrast to the treatment of Ref.~\cite{Gubser:2008as},
where the string energy was assumed to be well-described by
$E_{GGPR} \sim \sqrt{\lambda}/(\uo \sqrt{1{-}v^2})$, 
the energy of the strings considered in this paper
scale as $\sqrt{\lambda} \, u_h^2 /u^3_*$
and the strings themselves are approximately null --- this latter
fact completely fixes the corresponding geodesic parameter
$\r$ in terms of the initial string profile via the equations of motion. 

Despite these differences, it may be of interest
to compare the penetration depth of a gluon estimated in
Ref.~\cite{Gubser:2008as} with the result for
a light quark found in this paper.
In doing so, it is natural to
replace $E_{*}/2 \rightarrow E_*$,
in Eq.~(\ref{ggprresult}) when converting from a folded string modeling
a gluon to an open string describing a quark.
With this change, one may simply compare $\mathcal C_{GGPR}$ 
to our measured value of $\C = 0.526$.
Our result is larger than the estimates of Ref.~\cite{Gubser:2008as}
by 30--50\,\%.  

The $E^{1/3}$ scaling of the penetration depth has also appeared in
Ref.~\cite{Hatta:2008tx},
which discussed the dynamics of jet-like configurations in the
bulk gauge field dual to the $R$-current in strongly coupled SYM.
The coefficient of the scaling relation was not calculated in this work.
However, the coefficient characterizing $R$-current jets
necessarily differs from our result in Eq.~(\ref{numFit}),
since the gravitational interactions of the $5d$ gauge field dual
to the boundary $R$-current are independent of $\lambda$
(at leading order in the strong coupling limit).

We conclude our discussion by summarizing the physics which 
distinguishes light quark energy loss from that of heavy quarks.  
A comprehensive numerical study of heavy quark evolution has been
performed in Ref.~\cite{Chernicoff:2008sa}.
Let us compare and contrast the behavior of heavy and light quarks.
The penetration distance in both cases is
non-universal for the same reason: the quark's evolution
depends upon the initial gauge field.  After
several units of inverse temperature, the dual string in either
case becomes well approximated by small fluctuations on top of an analytic
solution.
As long as the quark is ultra-relativistic (regardless of its mass),
the appropriate analytic string solutions are null strings,
and the energy flux flowing down the string
is entirely determined by the non-universal small fluctuations. 
However, when a heavy quark has lost a sufficient amount of energy,
its dual string profile will be well approximated by
the non-null $v < c$ solutions obtained in Ref.~\cite{Herzog:2006gh}.
Thereafter, the heavy quark energy loss rate will be insensitive to 
fluctuations away from the analytic string profile,
and the energy loss rate will simply be proportional to the quark's momentum.
In contrast, the light quark energy loss rate remains sensitively
dependent on fluctuations during its entire trajectory.
This is a consequence of the fact massless quarks are always
ultra-relativistic, so their dual string
profile is nearly null at all times.
Consequently, the energy loss rate profile of a
light quark remains sensitive to the initial conditions
for an arbitrarily long period until thermalization.

\section{Conclusions}
\label{conclusions}

Using gauge/gravity duality, we have studied the penetration depth of
an energetic light quark moving through a strongly coupled $\Nfour$ SYM plasma.
An analytic asymptotic analysis shows that, for quarks which travel long
distances through the plasma, the worldsheet of the dual string description
nearly coincides with that of null string.
Both the analytic analysis, and explicit numerical computations, show that
for a given quark energy $E$, the maximum penetration depth
$\Deltax(E)$
scales as $E^{1/3}$.
Based on numerical results from a wide sampling of initial conditions,
we find 
$\Deltax(E) = (\C/T) \, (E/T \sqrt\lambda)^{{1}/{3}}$
with $\C \approx 0.5$.  We also find that the instantaneous energy loss rate 
of a light quark is not universal.
However, independent of initial conditions, we find that
the energy loss rate grows rapidly as the thermalization time is approached.
Consequently,
the thermalization of light quarks in strongly coupled $\Nfour$
super Yang-Mills
ends with an ``explosive'' burst of energy transfer to the plasma.

\begin{acknowledgments}
We thank J. Casalderrey-Solana, E. Iancu, H. Liu, G. Moore and D. Teaney
for useful comments and discussions.
This work was supported in part by the U.S. Department of Energy under
Grant No.~DE-FG02-96ER40956.
P.C. and L.G.Y.  thank the Kavli Institute for Theoretical Physics for its
hospitality during the completion of key parts of this paper.

\end{acknowledgments}

\bibliographystyle{utphys}
\bibliography{refs}

\end{document}